\documentclass[useAMS,usenatbib,fleqn]{mn2e}

\usepackage{aas_macros}

\usepackage{epsfig}
\usepackage{epstopdf}
\usepackage{subfigure}
\usepackage{amsmath}    
\usepackage{longtable}
\usepackage{dcolumn}
\usepackage{amssymb}

\usepackage{wrapfig}

\DeclareGraphicsExtensions{.jpg,.eps,.EPS,.png,.pdf,.ps}

\usepackage{colordvi}
\usepackage{ulem}

\RequirePackage[colorlinks=true
,urlcolor=blue
,anchorcolor=blue
,citecolor=blue
,filecolor=blue
,linkcolor=blue
,menucolor=blue
,pagecolor=blue
,linktocpage=true
,pdfproducer=medialab
]{hyperref}

\title{Radio Flares of Compact Binary Mergers: the Effect of Non-Trivial Outflow Geometry}

\author[B. Margalit \& T. Piran]{Ben Margalit$^{1,2}$ \& Tsvi Piran$^{1}$ \\
$^{1}$Racah Institute of Physics, Hebrew University of Jerusalem, Israel \\
$^{2}$Physics Department, Columbia University, 538 West 120th Street New York, NY 10027}

\date{}

\begin{document}
\maketitle

\parskip 5pt
\begin{abstract}

The next  generation gravitational waves (GW)  detectors 
are most sensitive to GW emitted by compact (neutron star/black hole) binary mergers.  If one of those is a neutron star 
the  merger will  also emit electromagnetic radiation via three possible channels: Gamma-ray bursts and their (possibly orphan) afterglows \citep{Eichler1989},  Li-Paczynski Macronovae \citep{LiPaczynski1998} and  radio flares \citep{NakarPiran2011}. This accompanying electromagnetic radiation is vitally important in confirming the GW detections \citep{KochanekPiran1993}. It could also reveal a wealth of information regarding the merger and will  open a window towards multi-messenger astronomy. Identifying and characterizing these  counterparts is therefore  of utmost importance.  In this work we explore late time radio flares emitted by the dynamically ejected outflows. We build upon previous work and  consider the effect of the outflow's non-trivial geometry.  Using an approximate method we estimate the radio light-curves for several ejected matter distributions obtained  in numerical simulations. 
Our method provides an upper limit to the effect of non-sphericity. Together with the spherical estimates the resulting light curves bound the actual signal. 
 We find that while non-spherical geometries can in principle  lead to an enhanced emission, in most cases they result in an  increase in the timescale  compared with a corresponding spherical configuration. This would weaken somewhat these signals and  might decrease  the detection prospects.
\end{abstract}

\section{Introduction} \label{sec:Introduction}

Compact binary mergers (hereafter called simply mergers) have recently been the focus of extensive research in the efforts to detect gravitational waves (GW) emitted in these merger events \citep[see][for a recent review]{FaberRasio2012}. The decreasing orbital period of the binary pulsars  \cite[e.g.][]{HulseTaylor1975,Taylor1989,Kramer06,Weisberg2010} provides so far the best evidence for the existence of  GW. Still, as yet, GW have not been directly detected. While gravitational radiation should be produced in various different scenarios, current GW detectors are most sensitive to radiation emitted in mergers, and hence the quest of detecting GW is closely linked with an understanding of these events. 

A detection of electromagnetic counterparts to these merger events is of utmost importance \citep{KochanekPiran1993}.  
Such electromagnetic signals, if observed, could help pave the way to confirm the first GW detection and would reveal a wealth of information about the merger process \citep[e.g.][]{NakarPiran2011}. Since mergers are considered as likely sources of short Gamma-Ray Bursts (GRBs) \citep{Eichler1989}, these have often been considered as possible electromagnetic counterparts to mergers. However GRBs are highly beamed and are only observed when their relativistic jet points towards us. This decreases significantly the chances of a coincident GW-GRB detection. Although off-axis GRBs cannot be observed in gamma-rays, they are followed by a late time isotropic radio signal, namely the orphan afterglow, which may possibly be observed. These orphan afterglows are more promising electromagnetic counterparts in this respect, yet their signals are expected to be weaker than the radio flares from dynamical ejecta from the mergers that we discuss here \citep{NakarPiran2011,HotokezakaPiran15}.

In addition to GRBs, two main electromagnetic counterparts have been suggested. These are the ``Macronova" (or alternatively ``Kilonova", as preferred by a few authors) which results from the radioactive decay of freshly synthesized r-process material and thus resembles a supernova \citep[][]{LiPaczynski1998, Kulkarni2005, Metzger2010, Piran2013, Kasen2013, BarnesKasen2013, TanakaHotokezaka2013,  Grossman2014}, and radio flares which arise from the interactions between mergers' outflows and the surrounding ISM and are analogous to Radio Supernovae and GRB afterglows \citep{NakarPiran2011,Piran2013}. The latter are produced via synchrotron radiation of non-thermal electrons accelerated at the outflow-ISM shock front. Owing to the typical velocities and masses involved in this ejecta ($\sim 0.1 c$ and $\sim 10^{-2} M_\odot$ respectively), this afterglow component should peak in the radio band on a timescale of a year after the merger. So far, radio flares have been considered only for spherically symmetric models \citep{NakarPiran2011,Piran2013}. Our goal here is to characterize the effect of departure from spherical symmetry on the resulting radio signals. This is an important refinement of previous results, since merger simulations suggests that the outflows produced by these mergers are significantly ashperical. For a related discussion on the effect of non-sphericity on the Macronova signal, see \cite{Grossman2014}.

In this work we treat only the sub/mildly relativistic outflow components ejected dynamically by mergers, and focus on the late time radio signal produced by these components. We neglect ultra-relativistic jet components that may produce GRBs or other short-lived signals, as well as non-dynamically ejected outflows such as those that may arise from strong neutrino driven winds \citep[see][for a review of these components and their expected radio flares]{HotokezakaPiran15}. 
Our method follows \cite{Piran2013}, hereafter PNR13, which assumed a spherically symmetric outflow, and address the impact of nontrivial geometry on the resulting radio signals and it's repercussions for the detectability of such signals using current and future radio telescopes.

The use of an  approximate method is justified in light of the large astrophysical uncertainties regarding these systems. In particular, the circum-binary density is a key unknown that will depend on the binary's location within  it's host galaxy. This density is crucial in reproducing the exact dynamics of the system, and variations in this density will affect the timescale and luminosity of the radio light-curves dramatically. Additionally, the exact outflows expected from mergers are still rather ill constrained. The dynamically ejected outflow depends on the exact merger scenario as well as on unknown microphysics and in particular on the neutron star's equation of state. We use the dynamical outflows found by \cite{Rosswog2014}, yet other groups using different numerical methods and equations of state find slightly different outflow characteristics \citep[e.g][]{Hotokezaka2013}. Furthermore, other mass outflows arise in mergers besides the dynamical ejecta considered here \citep[see e.g.][]{HotokezakaPiran15}. In view of these many uncertainties in the relevant physical parameters, we choose to use an approximate analysis instead of a full fledged numerical simulation. 
As we discuss later, our approximate method maximizes the impact of the non-sphericity and as such it provides an upper limit to the full effect.
Our motivation is to estimate the detectability of these late time radio signals in comparison with other suggested electromagnetic counterparts such as GRB orphan afterglows.
We therefore focus here on evaluating only the peak timescale and peak flux of these signals  (these will suffice for  estimating the detectability), and we do not presume to reproduce the exact radio flare light-curves. We expect that the real peak time scale and peak flux are somewhere between those estimtated here and those estimated assuming spherical symmetry.

This paper is organized as follows - we begin in \S\ \ref{sec:Spherical_Systems} by presenting the model for the dynamics and radiation of a spherical outflow. This model  follows and builds upon PNR13. In \S\ \ref{sec:Piecewise_Spherical} we present the piecewise-spherical approximation - our approximate method for treating aspherical outflows. We derive and discuss some general analytic results that are applicable to any system  in \S\ \ref{sec:Analytic_Estimates}. In \S\ \ref{sec:Numerical_Results} we apply our analysis to various dynamic ejecta configurations found by \cite{Rosswog2014}. We continue in \S\ \ref{sec:Detectability} by estimating the detectability prospects of these radio flares in light of the results presented in \S\ \ref{sec:Numerical_Results}. Finally, we conclude in \S\ \ref{sec:Discussion_and_Conclusions} and discuss the implications of the findings in the context of merger's electromagnetic counterpart searches. 

\section{Spherical Systems} \label{sec:Spherical_Systems}
Even though we are interested in non-spherical outflows in this work, our method is based  on results for spherical systems  (see \S \ref{sec:Piecewise_Spherical}). We begin, therefore, by  presenting and describing some basic features of radio-flares emitted by spherical systems.

The problem at hand, namely calculating the radio flare light curves expected from mergers' dynamical outflows, can be divided into two separate tasks. The first is calculating the hydrodynamics of the outflow, and in particular the dynamics of the outflow-ISM shock front. The second is calculating the synchrotron flux emitted at the shock front. We begin by recapitulating some essential results of PNR13 (including some basic equations that are needed for our formalism), and later discuss some additional properties of spherical outflows.

 An ejected shell of matter with energy $E$ and an initial velocity $v_0$  decelerates on a time scale, $t_{dec}$:
\begin{equation} \label{eq:t_dec}
t_{dec} = \frac{R_{dec}}{v_0} = \left( \frac{3E}{4 \pi n m_p } \right)^{1/3} v_0^{-5/3} ,
\end{equation}
where $n$ is the circum-binary ISM number density, and $m_p$ the proton mass.
The homologously expanding outflow  is characterized by the energy distribution $E\left( \geq v \right)$. At earlier times the ejecta  expands freely, whereas at later times the outflow follows the Sedov-Taylor self-similar solution \citep{Sedov1946,Taylor1950}.
At times between $t_{dec}$ of the fastest moving shell in the ejecta, and $t_{dec}$ of the slowest moving shell we describe the approximate dynamics of the contact discontinuity (between the shocked ejecta and shocked ISM) by solving:
\begin{equation} \label{eq:Dynamics_1}
\frac{4\pi}{3} n m_p R^3 v^2 = M(R)v^2 = E\left( \geq v \right) ,
\end{equation}
to obtain the radius $R(t)$ and velocity $v(t)$ of the contact discontinuity. The leftmost equality of Eq.\ \ref{eq:Dynamics_1} assumes spherical geometry and a constant ISM density. It is correct up to a factor of order unity, which will depend on the exact structure profile of the ejecta. The dynamics implied by this equation is governed solely by the outflow's angular energy density, $dE/d\Omega$, which is just $E/4\pi$ for the spherical case. In the following sections we generalize this for a non-spherical outflow confined to a solid angle $\Omega$, and we use this term ($dE/d\Omega$) here for convenience.

The collisionless shock wave accelerates electrons to a power-law distribution $dN/d\gamma \sim \gamma^{-p}$.  This holds for $\gamma > \gamma_m$, where $\gamma_m$ is the minimal Lorentz factor of the accelerated electrons, and $2 \leq p \leq 3$ is the power-law index which should be around $p \approx 2.1-2.5$ for mildly relativistic shocks, and around $p \approx 2.5-3$ for Newtonian shocks.
Fraction $\epsilon_e$ and $\epsilon_B$ of the total energy density at the shock front are given to  the electron and the magnetic energy respectively. 
Synchrotron radiation  is emitted by these accelerated electrons.    Following PNR13, the typical synchrotron frequency $\nu_m$ and typical flux $F_m$ can be written as:
\begin{equation} \label{eq:nu_m}
\nu_m(t) \approx 1.3~\text{GHz} \left(\frac{\epsilon_B}{0.1}\right)^{1/2} \left(\frac{\epsilon_e}{0.1}\right)^2 n^{1/2} \beta (t)^5 ,
\end{equation}
\begin{equation} \label{eq:F_m}
F_m(t) \approx 0.5~\text{mJy} \left(\frac{R(t)}{10^{17} \text{cm}}\right)^3 n^{3/2} \left(\frac{\epsilon_B}{0.1}\right)^{1/2} \left(\frac{d}{10^{27} \text{cm}}\right)^{-2} \beta(t) ,
\end{equation}
where $d$ is the distance between the source and the observer (for simplicity we neglecting any cosmological effects).

The synchrotron emission is self absorbed below $\nu_a$, the synchrotron self-absorption frequency.  For simplicity we focus on a single
 standard scenario in which $\nu_a,\nu_m < \nu_{obs}$,
where $\nu_{obs}$ is the observed frequency.  This generally holds for  $\nu_{obs} = 1.4~\text{GHz}$, a frequency for which  numerous facilities are available, as both $\nu_m$ and $\nu_a$ will normally be significantly smaller than $1~\text{GHz}$ (see Eq.\ \ref{eq:nu_m} and equation 5 of PNR13 plugging in typical velocities of the merger's dynamic ejecta, $\beta \approx 0.1-0.2$). In this case Eqs.\ \ref{eq:nu_m} and \ref{eq:F_m} are sufficient to determine the light curve. Since only the velocity shells which have already decelerated at time $t$ contribute to the signal, the light-curve scales as
\begin{equation} \label{eq:Flux_Scaling}
F_{\nu_{obs}}(t) = F_m(t) \left(\frac{\nu_{obs}}{\nu_m(t)}\right)^{-\frac{p-1}{2}} \propto E \left( \geq v(t) \right) v(t)^{-\frac{5p-7}{2}},
\end{equation}
where the last proportionality is given by relating $R(t)$ to the ejecta's energy distribution via Eq.\ \ref{eq:Dynamics_1}. It describes the fact that only the energy of  shells initially moving faster than $v(t)$ contribute. The complete light-curve is therefore given by solving Eq.\ \ref{eq:Dynamics_1} for the shock radius and velocity and substituting these into Eq.\ \ref{eq:Flux_Scaling}.

In the following we discuss some properties of light-curves arising from this model. In particular we examine the peak flux and peak timescale for these signals which are important for the rest of this work, but also quantify the shape of the light-curve by defining the peak width. We begin by examining an idealized toy model for the outflow energy distribution $E \left( \geq v \right)$, namely the single-velocity shell, and continue by showing that any arbitrary energy distribution can be approximated by this toy model for the purposes of evaluating the peak light-curve properties.

A spherical ``single-velocity shell" is a spherically symmetric outflow with energy distribution $dE/dv = E_0 \delta \left( v_0 \right)$. While this distribution does not well describe the merger's outflows, we will show later on that results for single-velocity shells can be very useful in approximating the peak light-curve properties (see \S\ \ref{sec:Analytic_Estimates}).

The light-curve of a single-velocity shell peaks at the deceleration timescale:\begin{equation}\label{eq:timescale}
t_{peak} = t_{dec} \propto \left(\frac{dM}{d\Omega}\right)^{5/6}\left(\frac{dE}{d\Omega}\right)^{-1/2} \ .
\end{equation}
The corresponding peak flux can similarly be expressed as:
\begin{equation}\label{eq:flux}
F_{peak} \propto E \left(\frac{dE}{dM}\right)^{-\frac{5p-7}{4}} \ .
\end{equation}

The  light-curve of a single-velocity shell scales as
\begin{equation} \label{eq:SingleVelocity_Flux}
F_{\nu_{obs}}(t) \propto \left\{
\begin{array}{lr}
t^3 & ; t \leq t_{dec} \\
t^{-\frac{15p-21}{10}} & ; t_{dec} < t
\end{array} \right. .
\end{equation}
Using this result we find the times at which the flux reaches half it's peak value ($t_{1/2}^{(-)}$ and $t_{1/2}^{(+)}$):
\begin{align} \label{eq:t-}
t_{1/2}^{(-)} = 2^{-1/3} t_{peak} \approx 0.8 ~t_{peak} ,
\end{align}
\begin{align} \label{eq:t+}
t_{1/2}^{(+)} = 2^{{10}/(15p-21)} t_{peak} \approx 1.5 ~t_{peak} ,
\end{align}
where in the last equality in Eq.\ \ref{eq:t+} we have substituted $p=2.5$. With  these expressions, we  define the peak width, $\Delta t$, as the time period during which the flux exceeds half it's peak value. For a spherical single-velocity shell, we find $\Delta t = t_{1/2}^{(+)}-t_{1/2}^{(-)} \approx 0.73 ~t_{peak}$. We use this measure for the peak width to quantify the extent with which two distinct light-curves overlap.

An important note regarding $\Delta t$ given above, is that the light-curve scaling laws on which it is based are self-similar solutions towards which (or from which) the system steadily evolves, and are only applicable for a single velocity shell. For more `realistic' shells which have an arbitrary $E\left( \geq v \right)$, the light-curve around the peak flux differs significantly from this approximation, changing smoothly between these scaling laws which only hold at very early times or late times. This effect significantly increases the peak width for non single-velocity shells. Still, $\Delta t$ is useful in that it may be used as a (rather loose) lower bound on the peak width of an outflow.

\begin{figure} \label{fig:E_v_distribution}
\centering
\epsfig{file=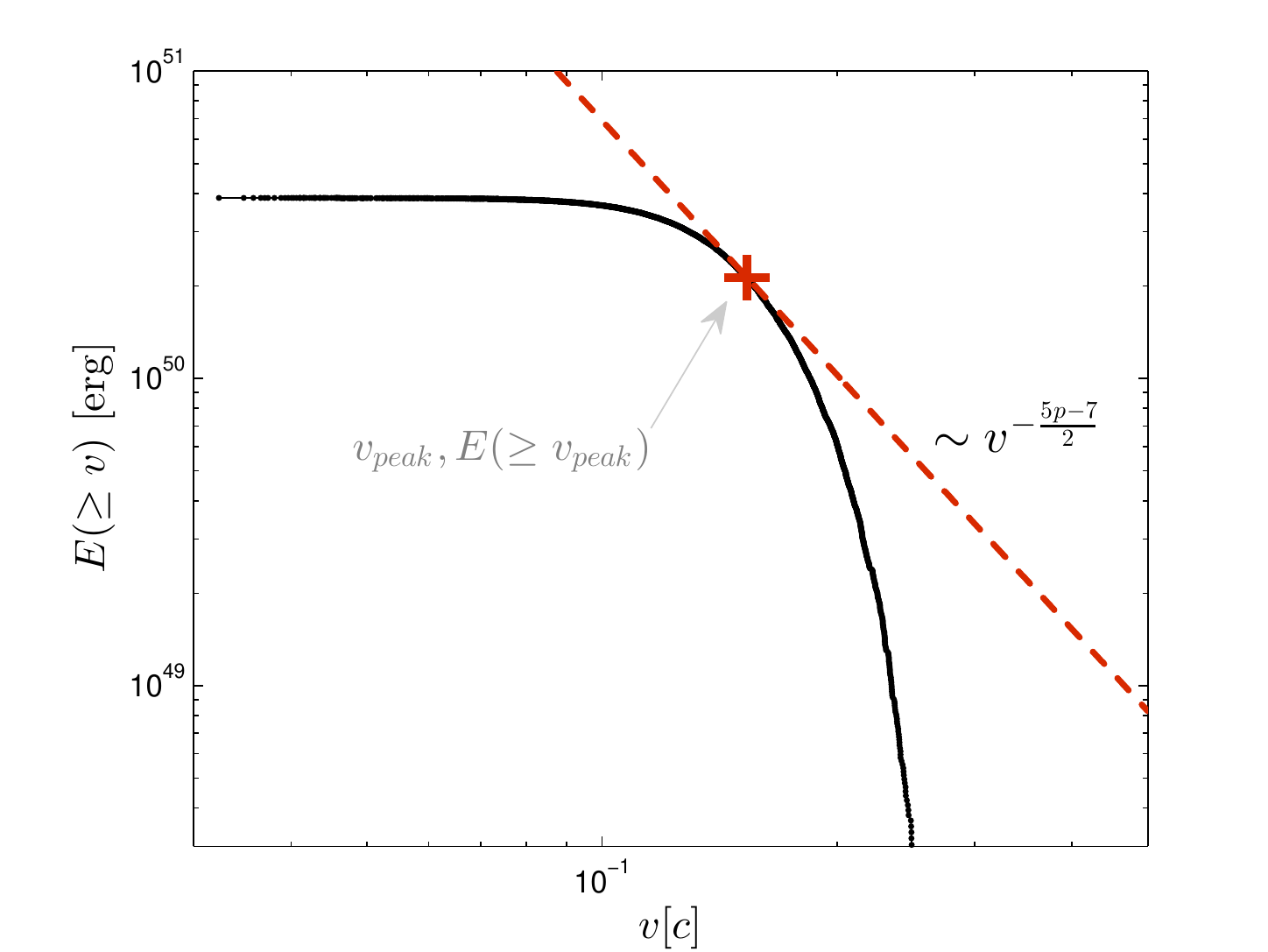,angle=0,width=0.5\textwidth}
\caption{Cumulative energy distribution $E(\geq v)$. The red cross signifies the velocity at which the calculated flux peaks. The dashed red line shows that the logarithmic slope at this peak velocity is $-(5p-7)/2$, as expected by Eq.\ \ref{eq_v_peak}. The cumulative distribution decreases faster than this slope above $v_{peak}$, but not as steeply as a step function (which would describe a single-velocity shell).}
\end{figure}

 We turn now to examine an arbitrary energy distribution $E \left( \geq v \right)$ outflow. The velocity $v_{peak}$ at which the light-curve peaks for such an outflow is easily found by differentiating Eq.\ \ref{eq:Flux_Scaling} with respect to $v$. 
This peak velocity satisfies:
\begin{equation}\label{eq_v_peak}
\left. \dfrac{d \mathrm{log} E(\geq v)}{d \mathrm{log} v} \right\vert_{v_{peak}} = -\frac{5p-7}{2} \ .
\end{equation}
This result is straightforward once Eq.\ \ref{eq:Flux_Scaling} is established - at earlier times than $t_{peak}$ the shock velocity is larger than $v_{peak}$, yet the amount of energy available $E(\geq v)$ decreases faster than $v^{-(5p-7)/2}$, hence the overall flux is smaller. At later times, the shock velocity is smaller than $v_{peak}$ and the energy $E(\geq v)$ increases, yet this increase is slower than $v^{-(5p-7)/2}$, hence the decrease in velocity dominates and the total flux is once again smaller. Figure \ref{fig:E_v_distribution} depicts a representative energy distribution for one of the merger's dynamic outflow configurations discussed below (ns14ns14), and shows that the velocity $v_{peak}$ at which the light-curve peaks corresponds to the value expected from Eq.\ \ref{eq_v_peak}.

We find that the peak flux of a radially structured outflow can therefore be characterized using only two parameters $v_{peak}$, and $E(\geq v_{peak})$. It  does not depend on the entire distribution $E(\geq v)$. Unfortunately we cannot characterize the time at which the light-curve peaks using only these two variables, for that the detailed energy distribution must be accounted for. Still, we may find $t_{peak}$ up to a factor of order unity using only these two parameters.

We can change the time at which the $v_{peak}$ velocity shell decelerates while keeping $E(\geq v_{peak})$ fixed by allocating more/less energy into higher velocity shells. If a large fraction of $E(\geq v_{peak})$ is put into very fast shells (which decelerate quickly), only a small fraction is left in the $v_{peak}$ shell, and by Eq.\ \ref{eq:t_dec}, this velocity shell decelerates earlier. An upper bound on this deceleration time is therefore given by the limiting case in which all the matter above this velocity, $M(\geq v_{peak})$, moves with $v_{peak}$ (as a single-velocity shell), and the distribution $E(\geq v)$ decreases infinitely fast above $v_{peak}$, as a step function. In reality the distribution will decrease gradually, some of the energy will go into faster shells with short deceleration times, and the  shell moving with $v_{peak}$  will be affected earlier. Thus this approximation will be altered by a numerical factor $<1$ but of order unity, that depends on the exact energy distribution. The peak time is bound from above by
\begin{equation} \label{eq:peakTime_upperbound}
t_{peak} \lesssim t_{dec} = \left( \frac{3E(\geq v_{peak})}{4\pi n m_p} \right)^{1/3}v_{peak}^{-5/3} \ .
\end{equation}

A lower bound on this numerical factor (and thus on the timescale) may be obtained by considering the case where $E(\geq v)$ falls off as slowly as possible above $v_{peak}$. Eq.\ \ref{eq_v_peak} in conjunction with the fact that both $E(\geq v)$ and $d \mathrm{log}E/d \mathrm{log}v$  decrease monotonically, implies that $E(\geq v)$ must falloff steeper than $v^{-(5p-7)/2}$. Hence we may find the peak time in the limiting case where $E(\geq v)$ decays as a power law above $v_{peak}$ with a slope $-(5p-7)/2$. By solving Eq.\ \ref{eq:Dynamics_1} in case of a general power law profile $E(\geq v) \propto v^{-(k-5)}$, we can write the expression for the shock velocity (as given by equation 16 of PNR13). We are interested in the case of $k=(5p+3)/2$. The peak time is thus bound from below by the time at which $v(t)$ for this power law profile equals $v_{peak}$. Solving this condition by substituting the appropriate $k$ and replacing $v_{min}$ by $v_{peak}$ in equation 16 of PNR13, we find that
\begin{equation} \label{eq:peakTime_lowerbound}
t_{peak} > \frac{k-3}{k} \left( \frac{3E(\geq v_{peak})}{4\pi n m_p} \right)^{1/3}v_{peak}^{-5/3} = \frac{5p-3}{5p+3} t_{dec} \ .
\end{equation}

Since both the upper and lower bounds scale equally (as $t_{dec}$), we have pinpointed the peak time to within a factor of $(5p-3)/(5p+3) \approx 0.6$. We can therefore reasonably characterize a complicated outflow distribution by a characteristic velocity $v_{peak}$ (which is defined by Eq.\ \ref{eq_v_peak}), and a characteristic energy $E(\geq v_{peak})$, and reduce realistic radially structured ejecta into roughly equivalent single-velocity shell building blocks. These `equivalent' single-velocity shells (characterized by velocity $v_{peak}$ and energy $E(\geq v_{peak})$) will not reproduce identical light-curves, but will reproduce the correct peak flux, and approximately the correct peak time. Since these are the focus of this present work, such a reduction will suffice when necessary.

\section{The Piecewise Spherical Approximation} \label{sec:Piecewise_Spherical}

So far we have discussed spherical outflows. Turning now to non-spherical outflows we  use a Piecewise Spherical Approximation in which
 we do not calculate the full three-dimensional (3D) hydrodynamics of the system in detail. Instead, we decompose the 3D system into many effective one-dimensional (1D) radial problems which are  solved separately. We then combine the one dimensional results to estimate the overall outflow evolution and the corresponding synchrotron emission.

The method follows schematically the following stages:
(i) We divide the system into  solid angle elements and extract the cumulative energy distribution $E\left(\geq v \right)/\Omega$ for each element.
This distribution determines the dynamics of each  element.
(ii) We calculate the spherical equivalent dynamics  of each solid angle using $E\left(\geq v \right)/\Omega$ according to Eq.\ \ref{eq:Dynamics_1}.
(iii) We obtain the resulting synchrotron  flux for each element. 
(iv) We estimate the system's overall light-curve  by summing the contribution of all elements.

We call this approximation ``piecewise-spherical" in the sense that we decompose the aspherical outflow into several solid angle `pieces' (denoted by index $i$) and treat each solid angle component $i$ as if it is a part of a spherically symmetric outflow with an energy distribution $dE(\geq v)/d\Omega = E_i(\geq v)/\Omega_i$. Using the results of \S\ \ref{sec:Spherical_Systems} for spherically symmetric outflows, we have a clear prescription of how to obtain the dynamics and synchrotron flux emitted by each component's spherical equivalent system, and summing these together yields the overall light-curve under this approximation. The crux of the approximation is the fact that we treat each solid angle component separately and independently from it's surrounding ejecta components. This simplifies enormously the dynamics and the calculations but it has its pitfalls. In the following subsection we describe the limitations of this approximate method.

\begin{figure*}
\centering
\epsfig{file=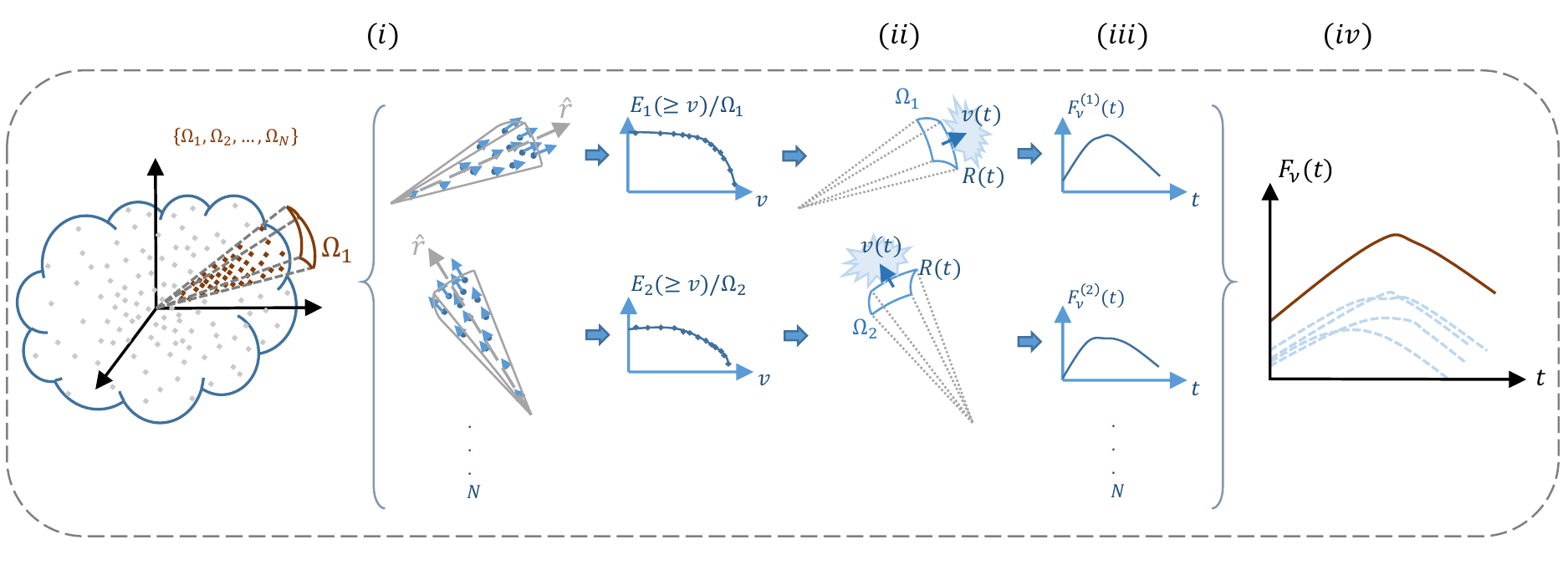,angle=0,width=0.95\textwidth}
\caption{A cartoon sketch illustrating our method. In step (i) we divide the spherical system into $N$ solid angle bins $\Omega_1,...,\Omega_{N}$, and extract the cumulative energy distribution $E(\geq v)$ from each. In step (ii) we use each bin's energy density distribution to calculate the temporal evolution of the spherical equivalent system via Eq.\ \ref{eq:Dynamics_1}. Using these dynamics, we calculate in 
(iii) the flux emitted by each solid angle. Finally we arrive at the total light-curve by summing up the $N$ contributions in (iv).}
\label{fig:Method}
\end{figure*}

\subsection{Limitations of the spherical-piecewise approximation} \label{subsec:Limitations_of_Our_Method}
We turn now to discuss the important issue of the limitations of this approximate method. We begin by stating clearly the underlying assumptions of the piecewise-spherical approximation:
\begin{enumerate}
\item[(1)] Each solid angle element evolves according to it's spherical equivalent system, and therefore the material in each solid angle expands only in the radial direction.
\item[(2)] Each solid angle element evolves separately, and is decoupled from any interactions with other elements. In particular, neighboring solid angles elements do not transfer mass or energy from one to another and they  don't  affect each other's dynamics.
\end{enumerate}

Both  assumptions will hold well if the angular variations in energy density and velocity are small, so that the system is nearly spherically symmetric. To illustrate this consider the limiting case of a jet confined within a solid angle $\Omega$. In this case the energy density and velocity change abruptly and dramatically on the boundary and assumption (1) will break down because the jet will expand azimuthally. This jet expansion occurs as the shocked region is heated and the hot shocked material expands into the surrounding cold ISM at roughly the speed of sound. If, on the other hand, we consider an angular region $\Omega$ which is a part of a nearly spherical system, such a sideways expansion is hindered.

Assumptions (1) and (2) are closely related to each other, as any azimuthal expansion of one bin essentially involves mass transfer to the adjacent bins. This mass transfer would in turn affect both bins' evolution and  assumption (2) will break down. However, we expect that even in case the system is far from spherical, both these assumptions will hold reasonably well {\it until the time of peak flux}. This is expected because the peak flux time in this regime ($\nu_m,\nu_a<\nu_{obs}$) is the deceleration time, $t_{dec}$, of the outflow. Because $t_{dec}$ is by definition the time at which the surrounding ISM begins affecting the ejecta's dynamics, it is reasonable to expect that the ISM-ejecta interactions will not result in any significant sideways expansion of the ejecta {\it before} this point.

Faster moving ejecta components will expand azimuthally before their slower surroundings, and energy will transfer from high to low velocity regions, effectively smoothing out the anisotropy in the initial distribution. Thus, our approximation overestimates the effect of asphericity, and we expect the piecewise-spherical method to provide an upper limit, such that the true light-curve lies somewhere in between our present results and those calculated using the spherical approximation.

\section{Analytic Estimates} \label{sec:Analytic_Estimates}

We begin by using the piecewise spherical approach described in \S\ \ref{sec:Piecewise_Spherical} to obtain simple analytic estimates that outline well the trends we observe later in the numerical simulations. To this end we compare simple 
 asymmetric outflows with their spherical equivalents, defined as having the same total energy and total mass. In particular we examine the effect of asymmetry on the peak time and peak flux of the radio signal. A useful picture in thinking about this problem, is that the transition from a spherical system into an aspherical one can be thought of as a ``redistribution" of the system's total mass and energy into various solid angle components. In other words, a spherical outflow with mass $M_{tot}$ and energy $E_{tot}$ uniformly distributed over a solid angle of $4\pi$ can be morphed into an aspherical outflow by redistributing the mass/energy unevenly between different solid angle elements. Under the piecewise spherical approach, each of these solid angle elements can then be evolved independently, allowing an estimate of the overall resulting light-curve.
To obtain an analytic intuition to the behavior of non-spherical systems we consider a simple model in which we approximate each solid angle element as a single-velocity shell, as opposed to treating the full  energy distributions $E(\geq v)$ for each solid angle element.

Consider  two simple examples: (a) The entire mass and the entire energy of the system are contained within half the sky ($2\pi~\text{rad}^2$). (b) The mass is distributed spherically but the entire energy is within $2\pi~\text{rad}^2$. In both cases, half the sphere contains no energy and therefore it does not contribute to the signal whatsoever. In case (a) the specific energy (and hence the velocity) of the relevant half sphere is the same as in the spherical case, yet the energy density has doubled. Therefore, the timescale increases by $2^{1/3}$, and the flux remains the same (the specific energy as well as the mass in half the sphere are identical to the spherical specific energy and total mass). On the other hand, in case (b) the relevant half spehere (the part which will contribute to the overall signal) contains the same mass density as in the spherically symmetric case, but twice the energy density. In this case the timescale decreases by $2^{-1/2}$, and the flux increases by $2^{(5p-7)/4}$.  This illustrates the fact that the timescale and the flux of an asymmetric system may either increase or decrease with respect to it's spherical equivalent depending on the details of the mass and energy redistribution.  

We now turn to apply the same considerations to a set of $N$ solid angle elements characterized by $dM_i, dE_i, d\Omega_i$ where the index $i$ enumerates the various solid angle elements. These parameters satisfy: $\sum_i dM_i=M_{tot}$, $\sum_i dE_i=E_{tot}$ and  $\sum_i d\Omega_i=4\pi$.
Since we are primarily interested in the total light curve's peak timescale, we can estimate this under the piecewise spherical approximation by averaging over the peak flux times of all constituent solid angle elements. The mean peak time is a sensible estimate for the overall peak timescale as long as the temporal density is large enough. In other words, if the constituent light-curves of each solid angle overlap significantly around their peak fluxes, this estimate will be valid. This can be quantified by stating that the peak times of all solid angle components should be within $\sim \Delta t$ of each other (see Eqs.\ \ref{eq:t-},\ref{eq:t+}). 

We define the {\it flux weighted} mean peak time normalized by the original peak time of the spherical equivalent system $t_{peak}^{(sphr)}$ as a dimensionless function $\tau$:
\begin{align} \label{eq:tau}
& \tau({\bf d\Omega}, {\bf dE}, {\bf dM}) = \frac{\left\langle t_{peak_i} \right\rangle}{t_{peak}^{(sphr)}} = \nonumber \\ 
& \frac{ \sum_{i=1}^{N} d\Omega_i^{-1/3} dE_i^{-1/2} dM_i^{5/6} \times \left( \frac{ dE_i^{\frac{5p-3}{4}} dM_i^{-\frac{5p-7}{4}} }{ \sum_{j=1}^{N} dE_j^{\frac{5p-3}{4}} dM_j^{-\frac{5p-7}{4}} } \right)}{(4\pi)^{-1/3} E_{tot}^{-1/2} M_{tot}^{5/6}} \nonumber \\
& = \frac{\sum \tilde{d\Omega_i}^{-1/3} \tilde{dE_i}^{\frac{5p-5}{4}} \tilde{dM_i}^{-\frac{15p-31}{12}} }{\sum \tilde{dE_j}^{\frac{5p-3}{4}} \tilde{dM_j}^{-\frac{5p-7}{4}} } ,
\end{align}
where in the last equality we have introduced the normalized variables $\tilde{x_i} \equiv x_i/x_{tot}$ (so that $0 < \tilde{x_i} < 1$ and $\sum \tilde{x_i} = 1$). This constraint eliminates three variables so that we are left with a total of $3 \times (N-1)$ free parameters.

\begin{figure*}
\centering
\begin{subfigure}[]
{\label{fig:timescaleContour_a} \epsfig{file=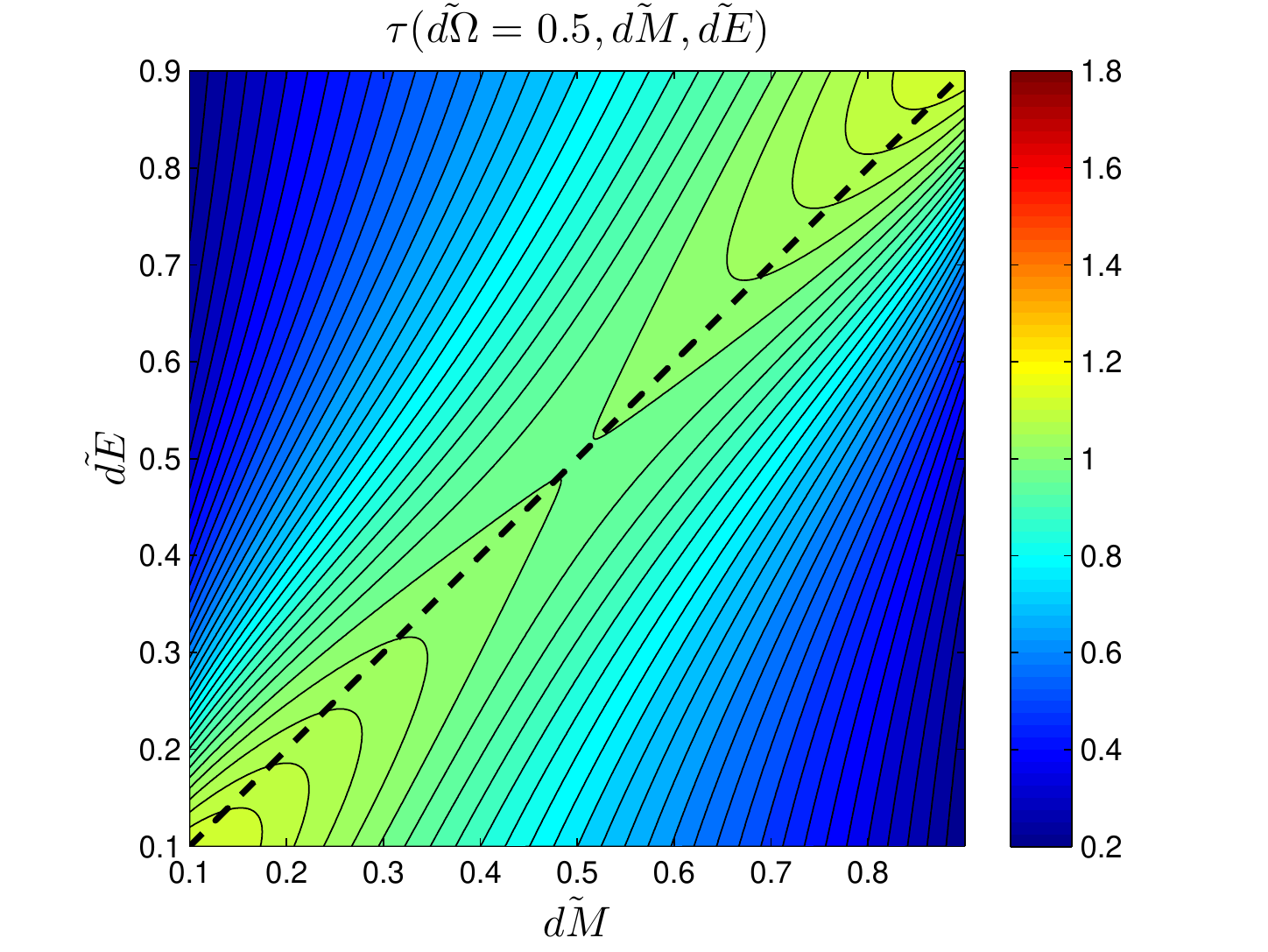,angle=0,width=0.45\textwidth} } 
\end{subfigure}
~
\begin{subfigure}[] 
{\label{fig:timescaleContour_b} \epsfig{file=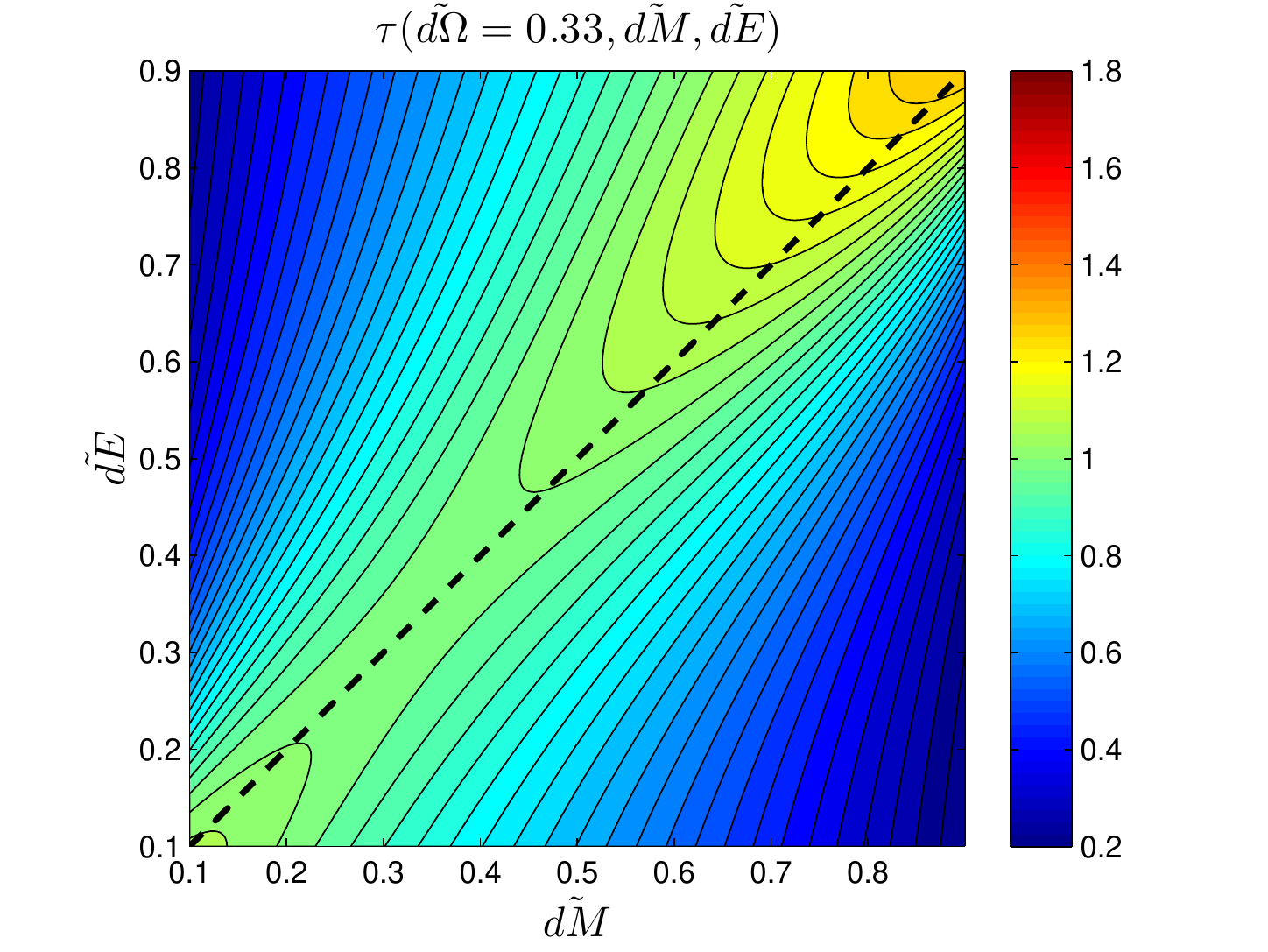,angle=0,width=0.45\textwidth} } 
\end{subfigure}
~
\begin{subfigure}[] 
{\label{fig:timescaleContour_c} \epsfig{file=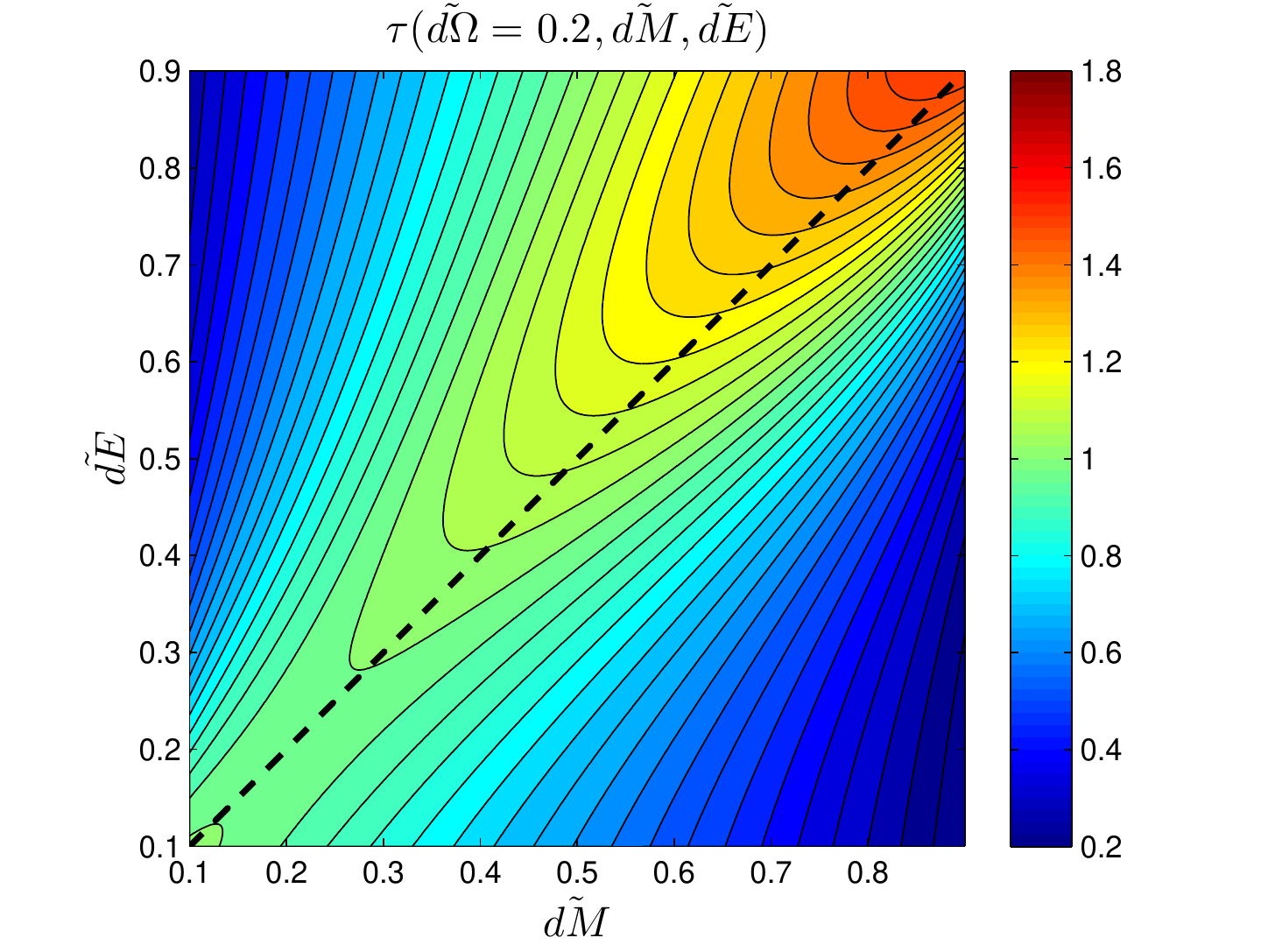,angle=0,width=0.45\textwidth} } 
\end{subfigure}
~
\begin{subfigure}[] 
{\label{fig:timescaleContour_d} \epsfig{file=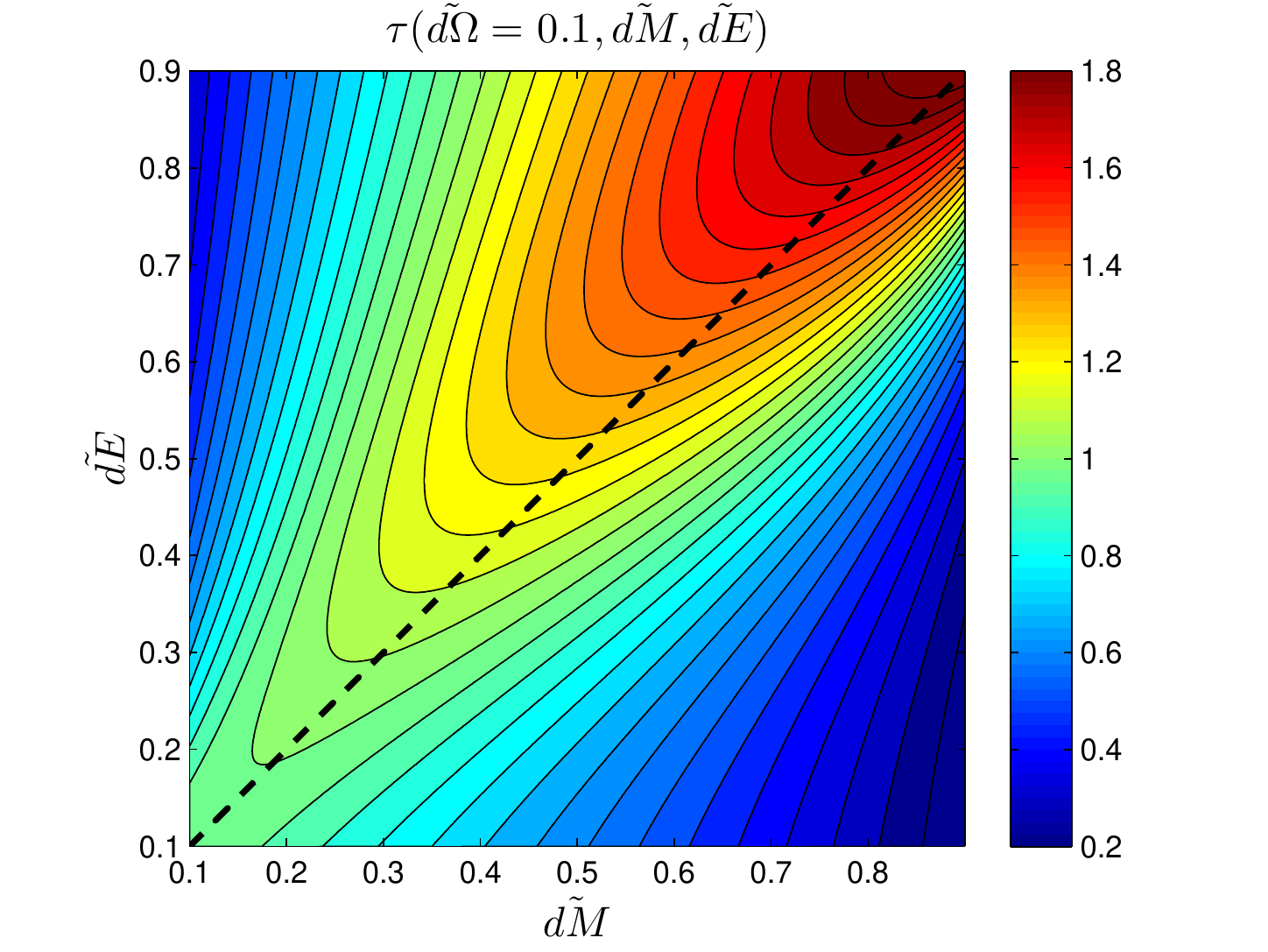,angle=0,width=0.45\textwidth} } 
\end{subfigure}
\caption{The weighted mean peak time (normalized with respect to the spherical peak time) - $\tau$, as a function of the redistribution variables $\tilde{d\Omega}, \tilde{dE}, \tilde{dM}$, for $N=2$ angular divisions. Plotted are contours of $\tau$ in the $(\tilde{dE}, \tilde{dM})$ plane, where successive figures (from top left to bottom right) depict decreasing values of $\tilde{d\Omega}$. The equal distribution point can be spotted in Fig. \ \ref{fig:timescaleContour_a} at $(\tilde{d\Omega},\tilde{dE},\tilde{dM})=(0.5,0.5,0.5)$ from which it is clear that this is a saddle point. Equal distribution points exist in the following plots as well (e.g. in Fig. \ \ref{fig:timescaleContour_c} the point is at $(\tilde{d\Omega},\tilde{dE},\tilde{dM})=(0.2,0.2,0.2)$), and the value of $\tau$ at these is always 1. The dashed black line corresponds to the constraint $\tilde{dE}/\tilde{dM}=1$, which physically means that the specific energy (and hence the velocity) is kept fixed. This line always passes through the equal distribution point, which serves as a minimum along this line. Therefore, with the additional constraint keeping $\tilde{dE}/\tilde{dM}$ fixed, the timescale necessarily increases ($\tau \geq 1$).}
\label{fig:timescaleContour}
\end{figure*}

Particularly interesting choices of parameters are ``equal distribution" points which retain the system's spherical symmetry (yielding $\tau=1$). Equal distribution points (e.d.) exist besides the trivial case of $\tilde{dx_k}=1/N$, as long as the mass and energy {\it densities} remain identical to the spherical distribution, i.e. $\tilde{dE_k}/\tilde{d\Omega_k}=\tilde{dM_k}/\tilde{d\Omega_k}=1$. These equal distribution points are critical points of $\tau$, since
\begin{equation}
\left. \frac{\partial \tau}{\tilde{\partial \Omega_k}} \right|_{e.d} = \left. \frac{\partial \tau}{\tilde{\partial E_k}} \right|_{e.d} = \left. \frac{\partial \tau}{\tilde{\partial M_k}} \right|_{e.d} = 0 .
\end{equation}
This extremum is a saddle point of $\tau$ and hence with appropriate choice of variables one can shorten or lengthen the timescale as desired ($\tau < 1$ and $1 < \tau$ respectively). We illustrate this fact graphically in Fig.\ \ref{fig:timescaleContour} by plotting contours of $\tau$ for the case of $N=2$.

We focus now on a particular scenario in which mass and energy are redistributed under the additional constraint that $dE/dM$ is kept constant. Under this constraint, all mass elements have equal velocity. In this case the equal distribution point discussed previously becomes a global minimum of $\tau$. The timescale of a constant specific energy redistribution can therefore only increase with respect to the spherical distribution: $\left. \tau \right|_{dE/dM=1} \geq 1$.

We show this fact graphically for $N=2$ in Fig.\ \ref{fig:timescaleContour}. The dashed black line in these figures depicts the curve $\tilde{dE}/\tilde{dM}=1$. Under the constant specific energy constraint, $\tau$ should be along this curve. It can be seen that the curve always passes through the equal distribution point at which $\tau=1$.  As one moves away from this point along the constraining curve (in either direction), the timescale only increases ($\tau > 1$).

An additional interesting point regarding the constant specific energy redistribution is that the total light-curve's peak flux in this case should remain roughly the same as its spherical equivalent. Assuming the underlying time dependent fluxes of each solid angle overlap sufficiently around their peak times (under the same conditions that render the mean peak time a reasonable estimate of the overall timescale), we may use Eq.\ \ref{eq:flux} to write the total light-curve's peak flux as the sum of it's components' peak fluxes, i.e.
\begin{equation}
\frac{F_{peak}}{F_{peak}^{(sphr)}} \lesssim  \sum_{i=1}^{N} \left(\frac{\tilde{dE}}{\tilde{dM}}\right)_i^{-\frac{5p-7}{4}} \tilde{dE_i} =  1 .
\end{equation}
Of course this is a rough order of magnitude estimate that can be used as an upper bound for $F_{peak}$, since the underlying peak fluxes should not perfectly overlap.

We have therefore shown that in case the specific energy is kept constant in a redistribution of mass and energy, the optically thin synchrotron peak time (as estimated by the weighted mean $\tau$) will necessarily increase, and the peak flux will not change with respect to the spherical equivalent light-curve. Of course one can always redistribute the mass and energy of the system creating a fast moving jet component. Such a jet will  peak at earlier times than the spherical equivalent system and may even lower the average peak time (so that $\tau<1$). This does not contradict our results since this type of redistribution does not obey the constraint of constant specific energy. In fact, such redistribution represents the exact opposite regime, in which there is a large (possibly extremely large) specific energy component and a second low specific energy component. Moreover, it is likely that in this case the weighted mean, $\tau$, breaks down as an estimate of the overall timescale, as the jet component's peak flux may exhibit itself as a completely separate component in the light-curve (in other words the two components will not significantly overlap). As long as the dynamic ejecta retains a roughly spherical distribution, one should not expect such a jet component or any other strong variations in specific energy. In this case our results stand and we expect a longer overall  light-curve timescale. In particular we will show in \S\ \ref{sec:Numerical_Results} that this is the case for the configurations we have studied and this serves as a  motivation in developing these results.

\section{Numerical Analysis of Calculated Ejecta Distributions} \label{sec:Numerical_Results}
We turn now to apply our method to the results of neutron star merger outflow simulations by \cite{Rosswog2014}, from which we obtain the 3D hydrodynamic configuration of the dynamical ejecta. These SPH simulations are a continuation of previous works \citep{Rosswog2013,Piran2013} which simulated neutron star mergers till several milliseconds after the merger event, and find that roughly $10^{-2} M_\odot$ is dynamically ejected from the mergers, with a strong dependence on any asymmetry in the binary masses (equal mass mergers do not tidally disrupt as much mass as non-equal binaries). \cite{Rosswog2014} followed up by simulating the long term evolution of the dynamical ejecta, taking into account the effects of nuclear heating on the outflow dynamics. The major hydrodynamic effect of this nuclear energy deposition is in accelerating slightly the outflow and diminishing significant irregularities in the outflow, essentially smoothing out the ejecta's angular distribution \citep[see][for more details]{Rosswog2014}. Since our calculations attempt to characterize the affects of asymmetry on the ejecta's afterglow signal, it is important we take the most realistic initial angular distribution as conditions for our calculations.

We  use snapshots of the dynamical ejecta, taken from \cite{Grossman2014} $\sim 10~\text{s}$ after the merger event, as initial conditions for our present calculations. We choose snapshots taken several seconds after the merger, as at these times the ejecta's interaction with the surrounding ISM is negligible and hence the calculations of \cite{Rosswog2014}, which did not account for these interactions, are  valid. On the other hand, these times are late enough so that the nuclear decays have deposited practically all their energy into the outflow and  affected it's dynamics. \cite{Rosswog2014} show that from this point onward the ejecta continues expanding homologeously (neglecting ISM interactions), and therefore the nuclear decays have little effect on the outflow at later times.

We examine outflows from simulations of three particular systems, distinguished by the neutron star masses: a canonical equal mass $1.4 M_\odot$-$1.4 M_\odot$ binary, a slightly asymmetric binary with masses $1.4 M_\odot$-$1.3 M_\odot$, and an extremely asymmetric system with masses $1.6 M_\odot$-$1.2 M_\odot$. We denote these systems `ns14ns14', `ns14ns13', and `ns16ns12' respectively. We additionally restrict our focus in this work to the case where $\nu_{obs}=1.4~\text{GHz}$ at which numerous radio facilities operate. This frequency is high enough so that we are always in the optically thin, $\nu_m,\nu_a < \nu_{obs}$, regime. Throughout this work we take the microphysical system parameters as: $\epsilon_e = 0.1$, $\epsilon_B = 0.1$, and $p=2.5$, and use an ISM number density of $n=1~\text{cm}^{-3}$, but these parameters are easily changed with qualitatively well known effects.

This section is divided into two subsections loosely corresponding to stages (i) and (ii-iv) of our method respectively (see Fig.\ \ref{fig:Method}). In \S\ \ref{subsec:Numerical_Results_subA} we discuss the division of the systems into angular bins, and present a reduction of the initial data (after binning) in terms of the characteristic energy and velocity of each bin. We continue in \S\ \ref{subsec:Numerical_Results_subB} by presenting the calculated light-curves  and discuss the results. Our main finding is an increase in the timescale of these light-curves in comparison to their spherical equivalents. This can be expected from the analytic estimates presented in \S\ \ref{sec:Analytic_Estimates}.

\subsection{Angular Binning} \label{subsec:Numerical_Results_subA}
Using spherical coordinates centered on the merger location, we divide each system's outflow into $N=20 \times 20$ solid angles, keeping the number of SPH particles in each solid angle fixed. In Appendix \ref{subsec:Errors} we show that this arbitrary binning allocation does not affect our results.

\begin{figure*}
\centering
\begin{subfigure}[ns16ns12 - characteristic energy density]{ \label{fig:ns16ns12_Mollweide} \epsfig{file=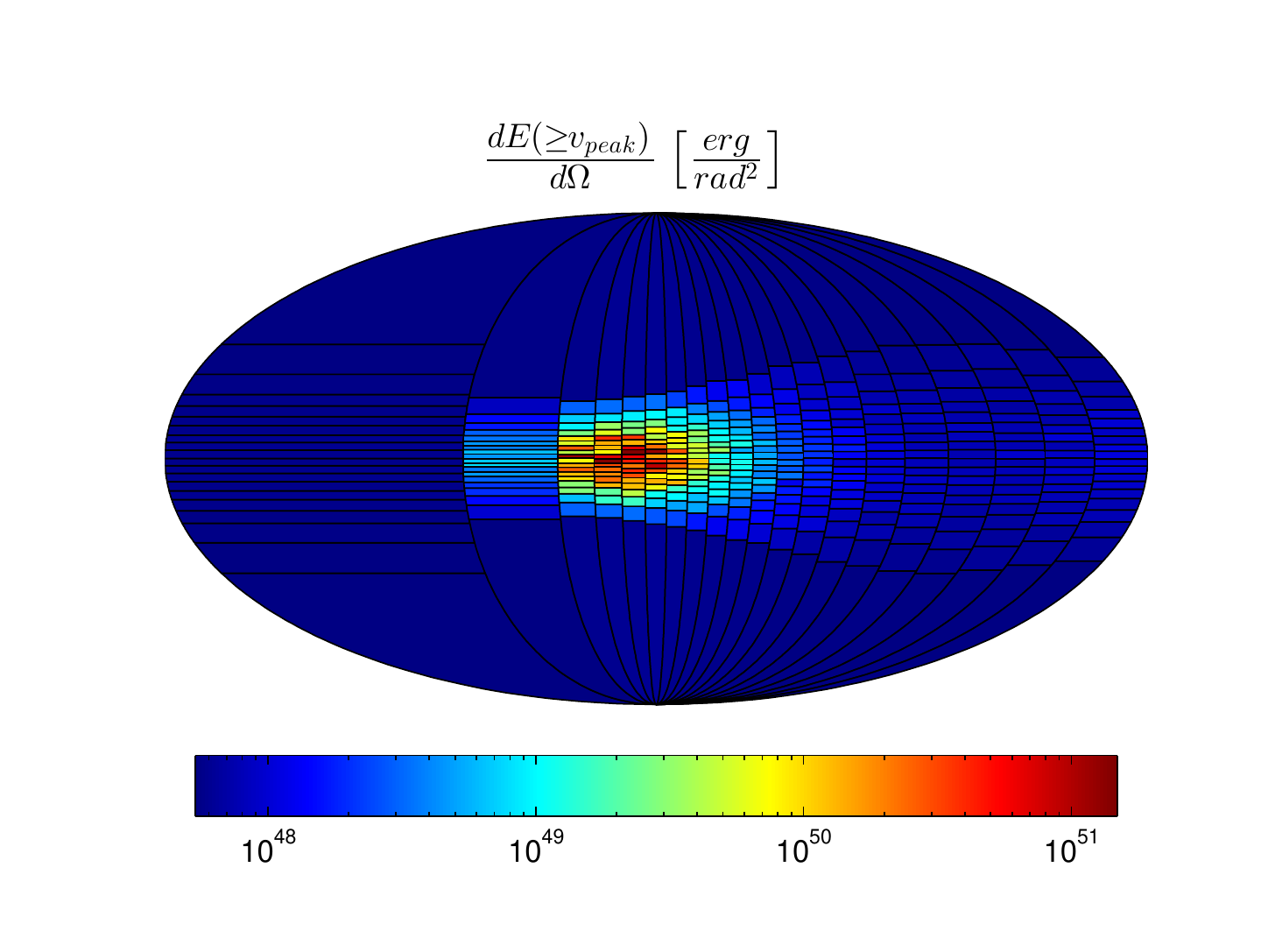,angle=0,width=0.45\textwidth} }
\end{subfigure}
~
\begin{subfigure}[ns16ns12 - characteristic velocity]{ \label{fig:ns16ns12_Mollweide_velocity} \epsfig{file=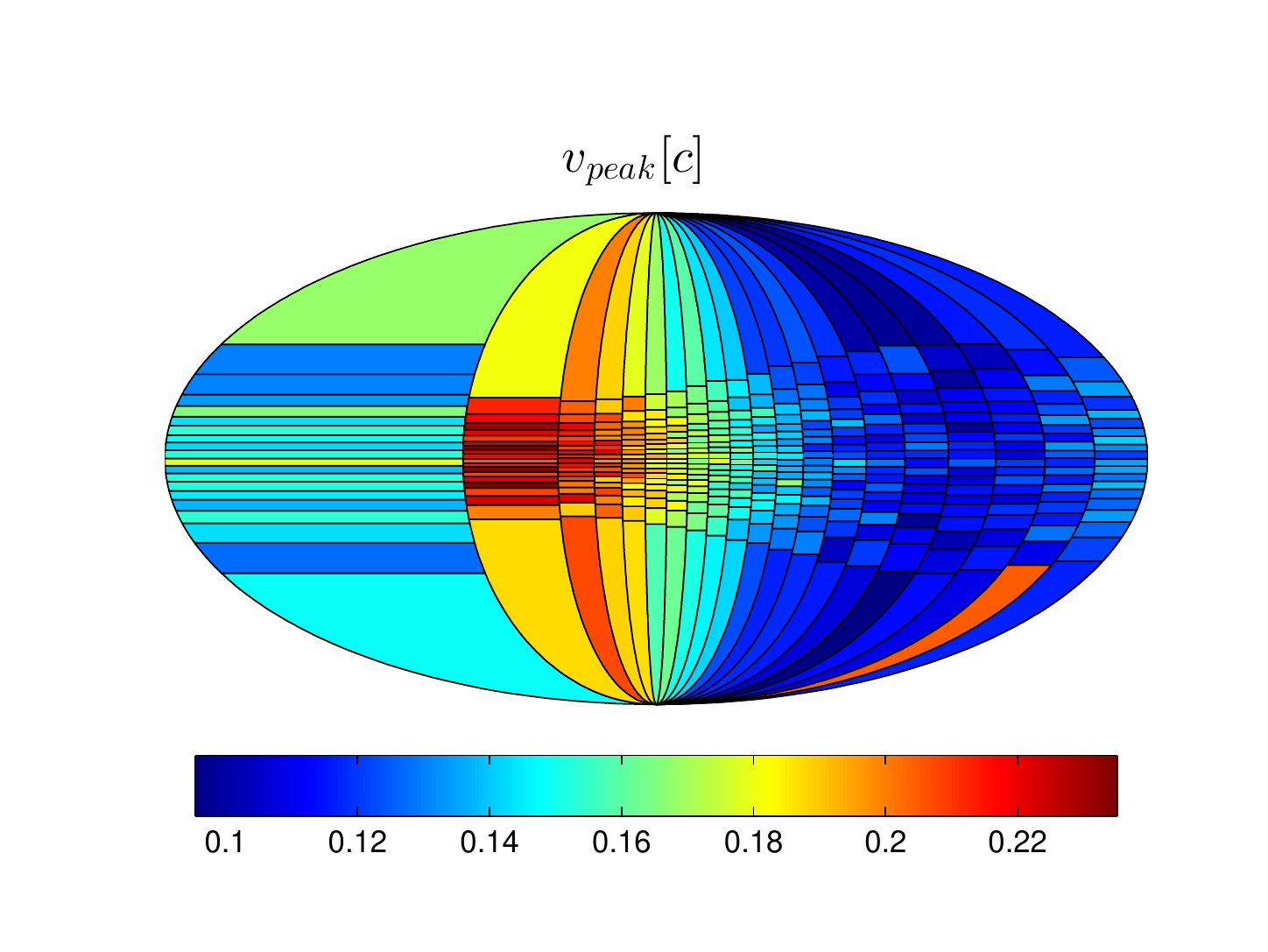,angle=0,width=0.45\textwidth} }
\end{subfigure}
~
\begin{subfigure}[ns14ns13 - characteristic energy density]{ \label{fig:ns14ns13_Mollweide}  \epsfig{file=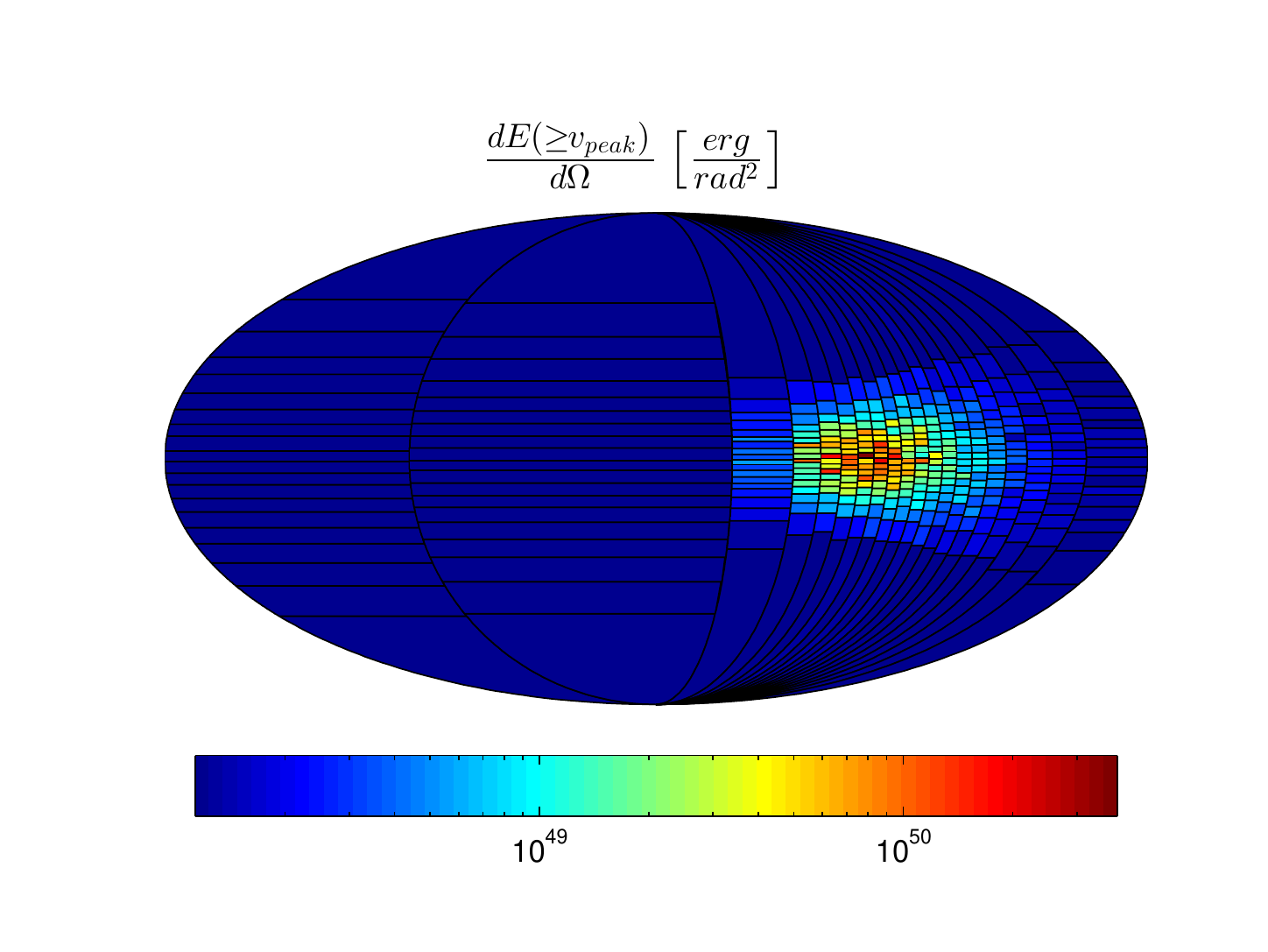,angle=0,width=0.45\textwidth} }
\end{subfigure}
~
\begin{subfigure}[ns14ns13 - characteristic velocity]{ \epsfig{file=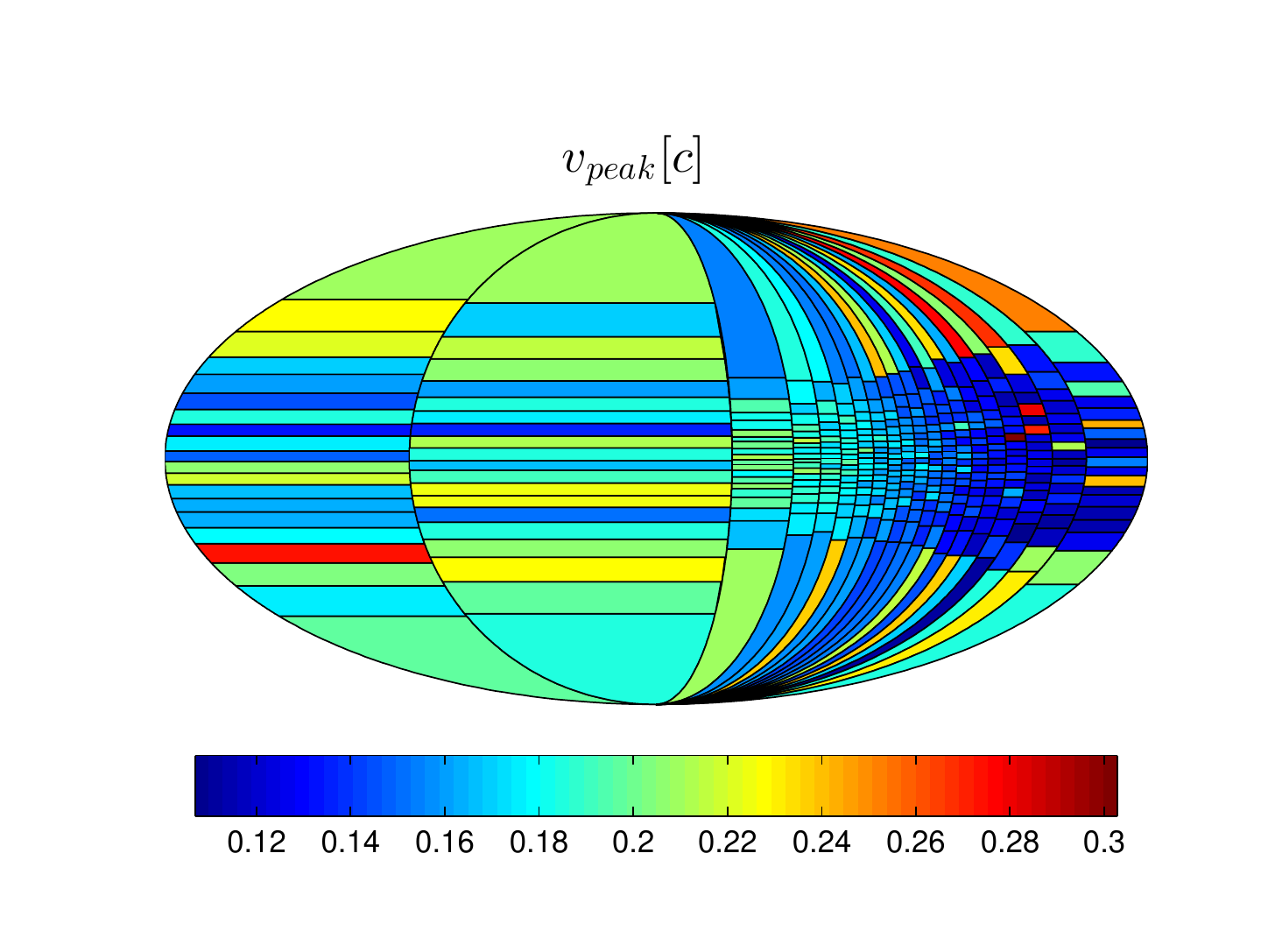,angle=0,width=0.45\textwidth} }
\end{subfigure}
~
\begin{subfigure}[ns14ns14 - characteristic energy density]{ \label{fig:ns14ns14_Mollweide}  \epsfig{file=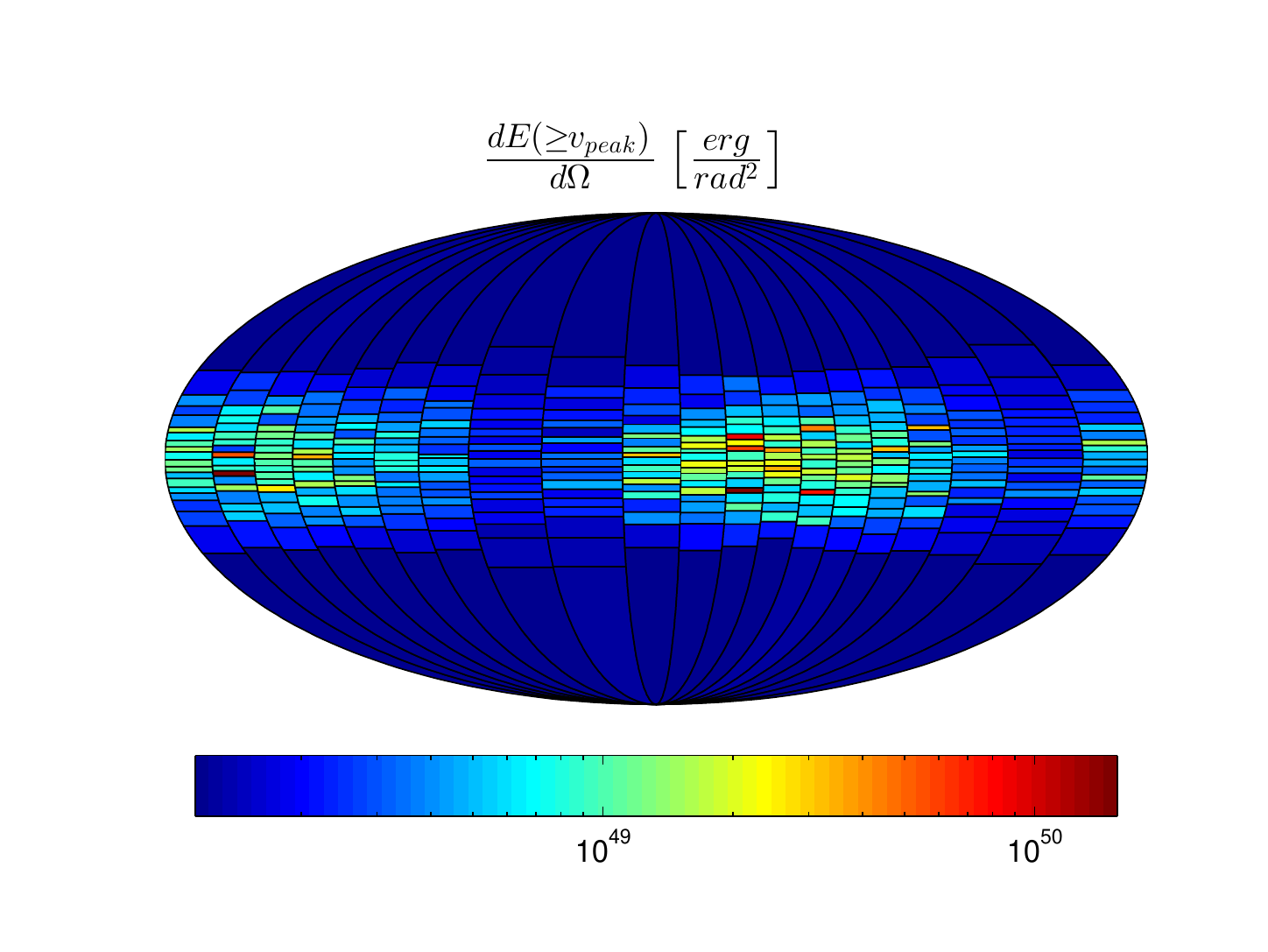,angle=0,width=0.45\textwidth} }
\end{subfigure}
~
\begin{subfigure}[ns14ns14 - characteristic velocity]{ \epsfig{file=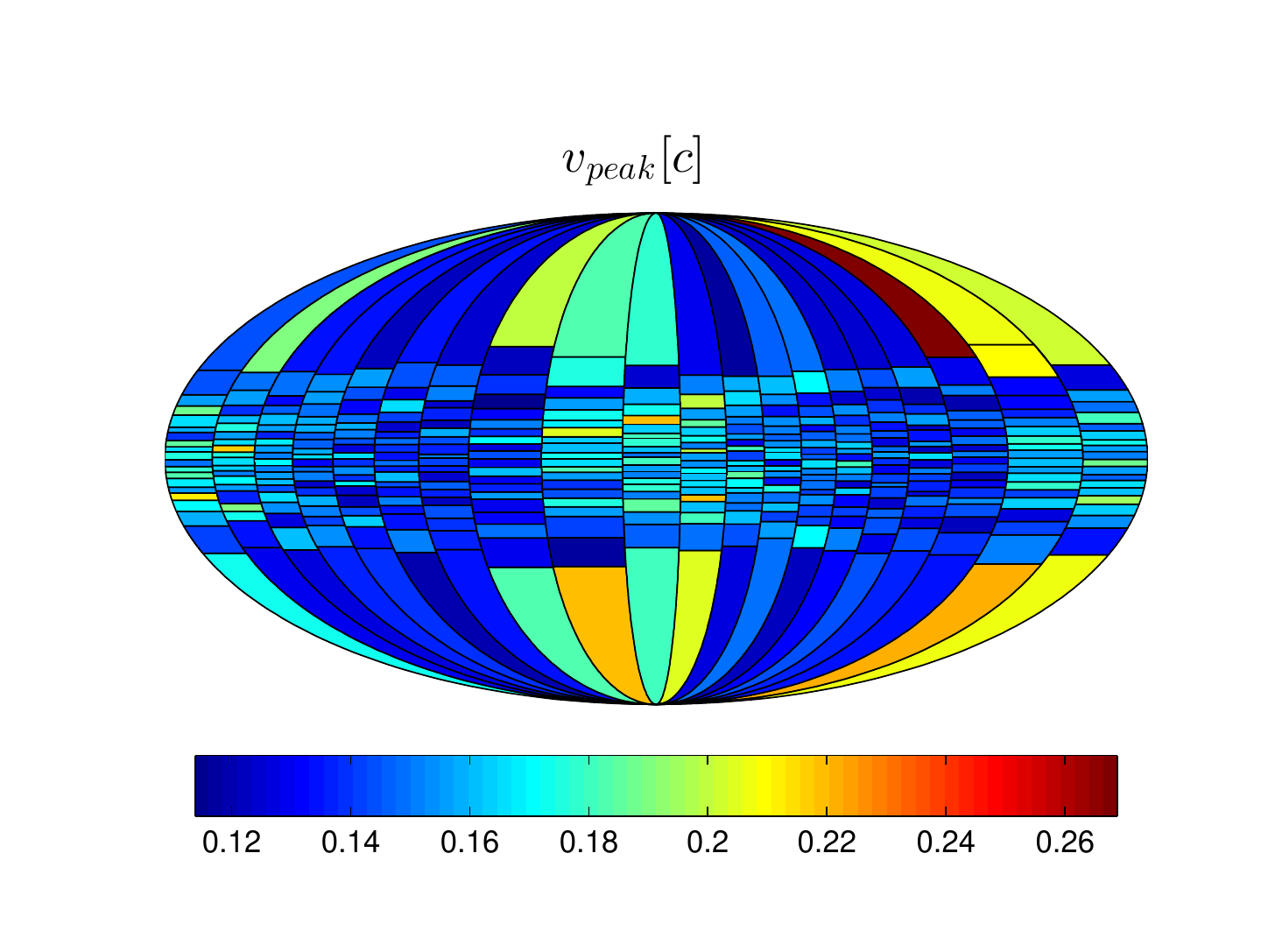,angle=0,width=0.45\textwidth} }
\end{subfigure}
\caption{Mollweide projections of the velocity and energy density angular distributions for ns16n12 (a,b), ns14ns13 (c,d) and ns14ns14 (e,f). The energy of the ejecta is concentrated in all three cases into a region close to the binary orbital plane ($\theta=0$). Additionally, the ejecta energy is concentrated in smaller $\varphi$ angles for larger binary mass ratios. This is a tracer of the tidally disrupted tail, which is  more prominent with increasing binary mass ratio. While the characteristic velocity varies by a modest factor of $\sim 2$, the energy density varies by nearly 3 orders of magnitude and it is therefore the decisive factor in determining the peak timescale.} \label{fig:Mollweide}
\end{figure*}

Figure \ref{fig:Mollweide} shows mollweide projections of the characteristic energy density $dE\left(\geq v_{peak}\right)/d\Omega$ and characteristic velocity $v_{peak}$ of each angular bin for the three  systems. These are a reduction of each bin's cumulative distribution $E(\geq v)$ into a single-velocity shell equivalent which reproduces the same peak flux and the approximate peak time (see \S\ \ref{sec:Spherical_Systems}).

The plots illustrate how the angular distributions tend toward symmetry as the progenitor merger masses become comparable. Still, even in the case of an equal masses merger, the distribution is far from spherical and there is a strong concentration around the merger plane (taken as $\theta=0$) so that the outflow is somewhat `flattened' \citep[see also Fig.\ 6 of][]{Rosswog2014}. The qualitative difference between equal and non-equal mass mergers is mostly apparent in that non-equal mergers show a stronger asymmetry inside the equatorial (merger) plane. This breakdown of the axial symmetry  is a trace of the tidally disrupted tails produced during the merger events. As expected unequal mass binaries  tidally disrupt each other more significantly than equal mass systems.

The azimuthal symmetry breaking is especially prominent in the case of the ns16ns12 merger shown in Fig.\ \ref{fig:ns16ns12_Mollweide_velocity}. Here a large jump in the characteristic velocity is apparent at $\varphi \approx 2\pi /3$, with no coinciding jump in the energy density. In fact, while the $\varphi \approx 2\pi /3$, $\theta \approx 0$ bins reach the largest velocities of the distribution, the same bins attain only moderate energy density values. The reason for this strange dichotomous behavior is that these bins catch on to the very end of the tidally disrupted tail ejected by the merger. This tail consists of a relatively low mass fast moving outflow, and hence insignificant energy density. As we decrease $\varphi$ further from this point, we find a large gap region containing no ejecta. This gap region is contained within the first ($0<\varphi\lesssim 2\pi /3$) bin, hence the sudden drop in $v_{peak}$. Therefore this unsmooth transition has it's roots in the asymmetry of the underlying equatorial distribution.

\begin{figure}[h]
\centering
\begin{subfigure}[]{ \label{fig:ns16ns12_equatorial_distribution} \epsfig{file=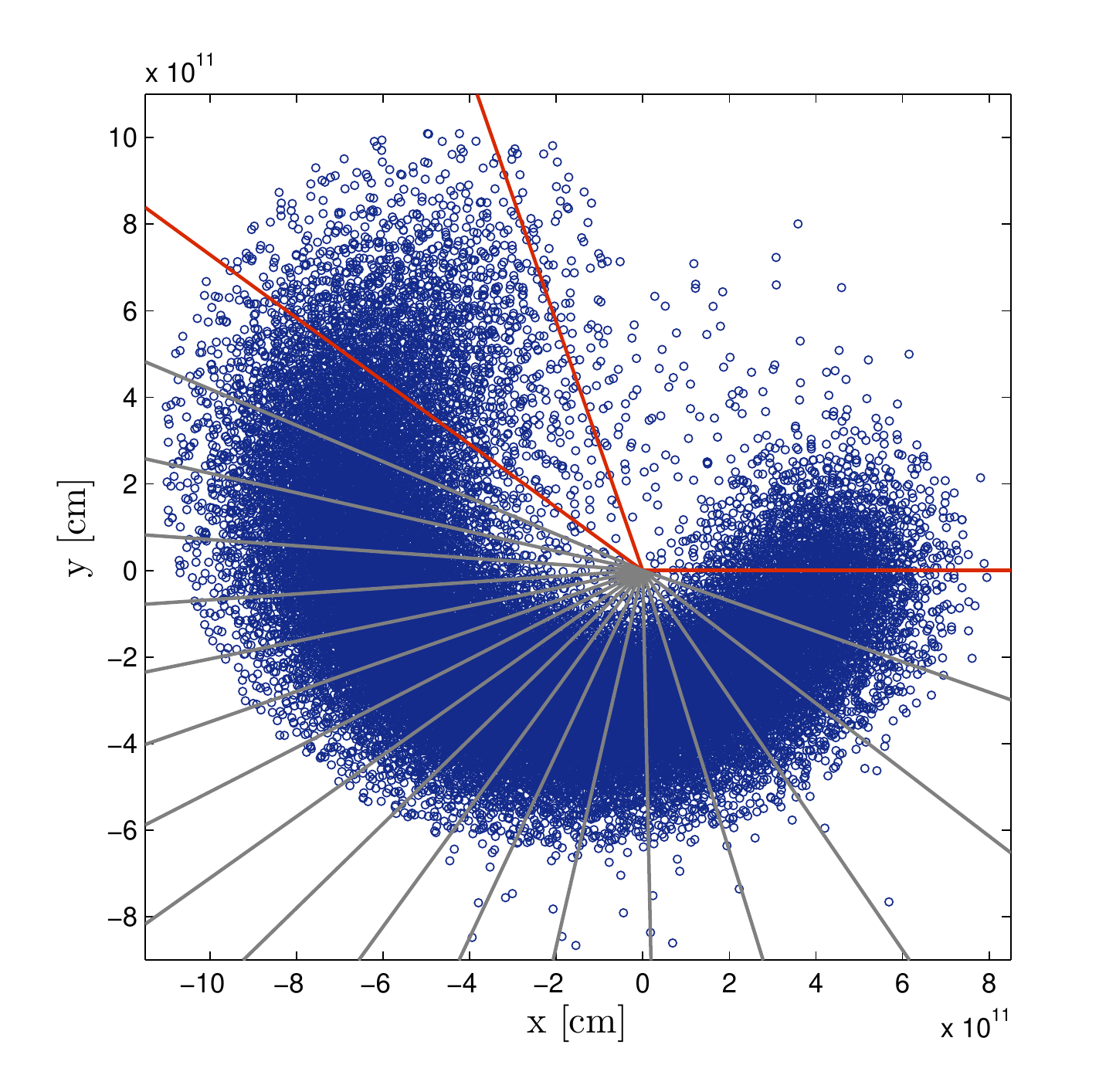,angle=0,width=0.45\textwidth} }
\end{subfigure}
~
\begin{subfigure}[]{ \label{fig:ns16ns12_equatorial_distribution_gap} \epsfig{file=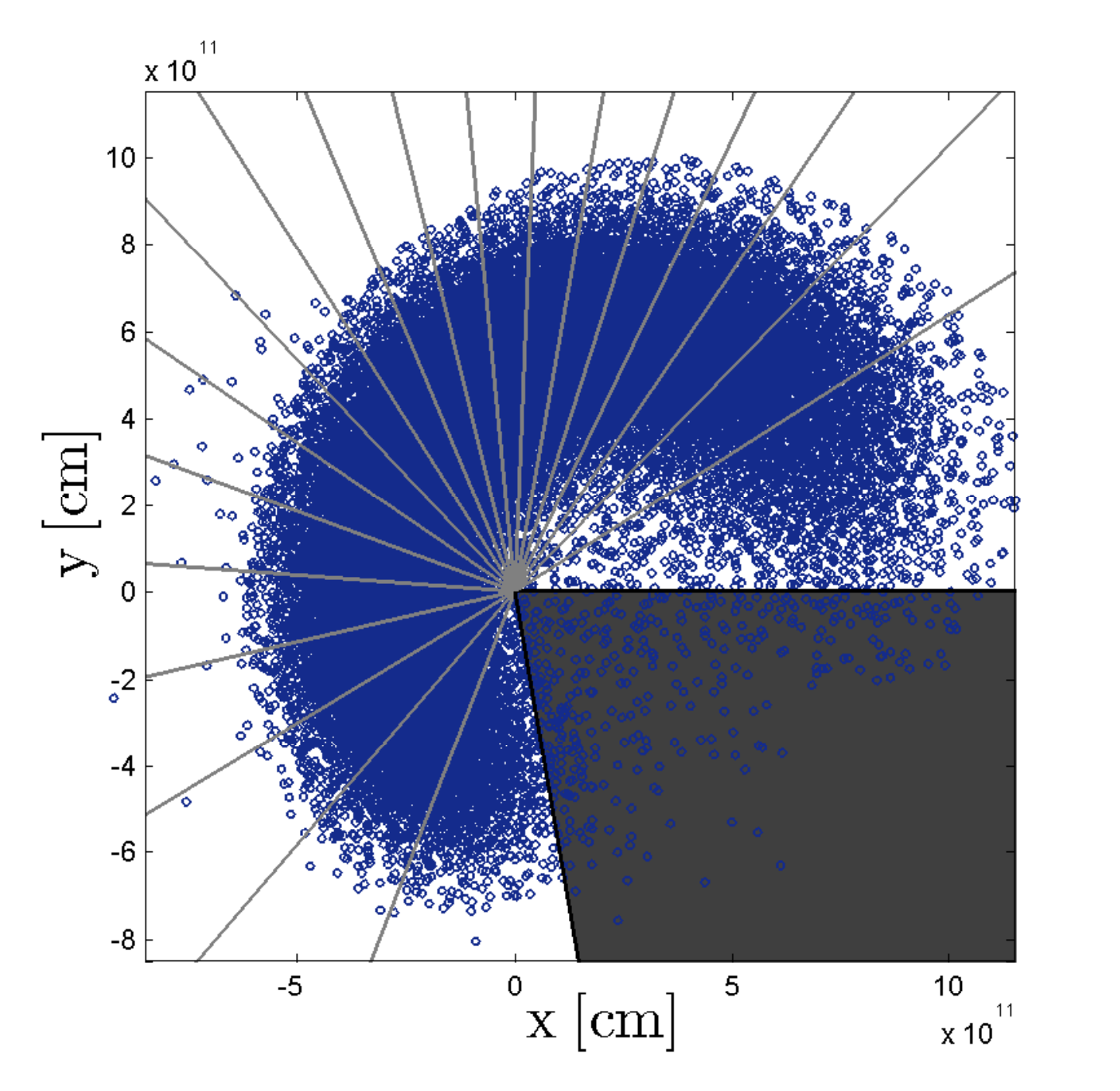,angle=0,width=0.45\textwidth} }
\end{subfigure}
\caption{The  SPH particle distribution for ns16ns12 projected on the equatorial (x-y) plane. Ejecta velocities are proportional to their radius. The radially extending grey lines correspond to the angular binning of the system in $\varphi$. (a) The original distribution. The red lines emphasize the bins between which a significant velocity jump occurs (see Fig. \ \ref{fig:ns16ns12_Mollweide_velocity}). This plot illustrates the fact that the velocity jump is due to the transition between the end of the tidally disrupted tail and a region containing slower moving ejecta. (b) The rotated distribution for the renewed analysis. The shaded black area indicates the gap region excluded from this analysis.}
\end{figure}

To illustrate this point further we plot the ns16ns12 SPH particle distribution projected onto the equatorial plane in Fig.\ \ref{fig:ns16ns12_equatorial_distribution}. The particles' velocities at this stage are very nearly proportional to their radii (as they are freely expanding) so that the distance  from the center is a tracer of the velocities. The angular binning in $\varphi$ is also plotted in this figure, allowing comparison with the mollweide projection of Fig.\ \ref{fig:ns16ns12_Mollweide_velocity}. The red lines indicate the two leftmost bins in Fig.\ \ref{fig:ns16ns12_Mollweide_velocity} over which the velocity jump occurs. This figure clearly shows that one bin is dominated by the end of the tidally disrupted tail, while the other bin is dominated by slower moving ejecta and encompasses the large gap region in the distribution (hence the significant angular size of this bin).

\begin{figure}
\centering
\begin{subfigure}[characteristic velocity map]{ \label{fig:GapRegion_mollweide} \epsfig{file=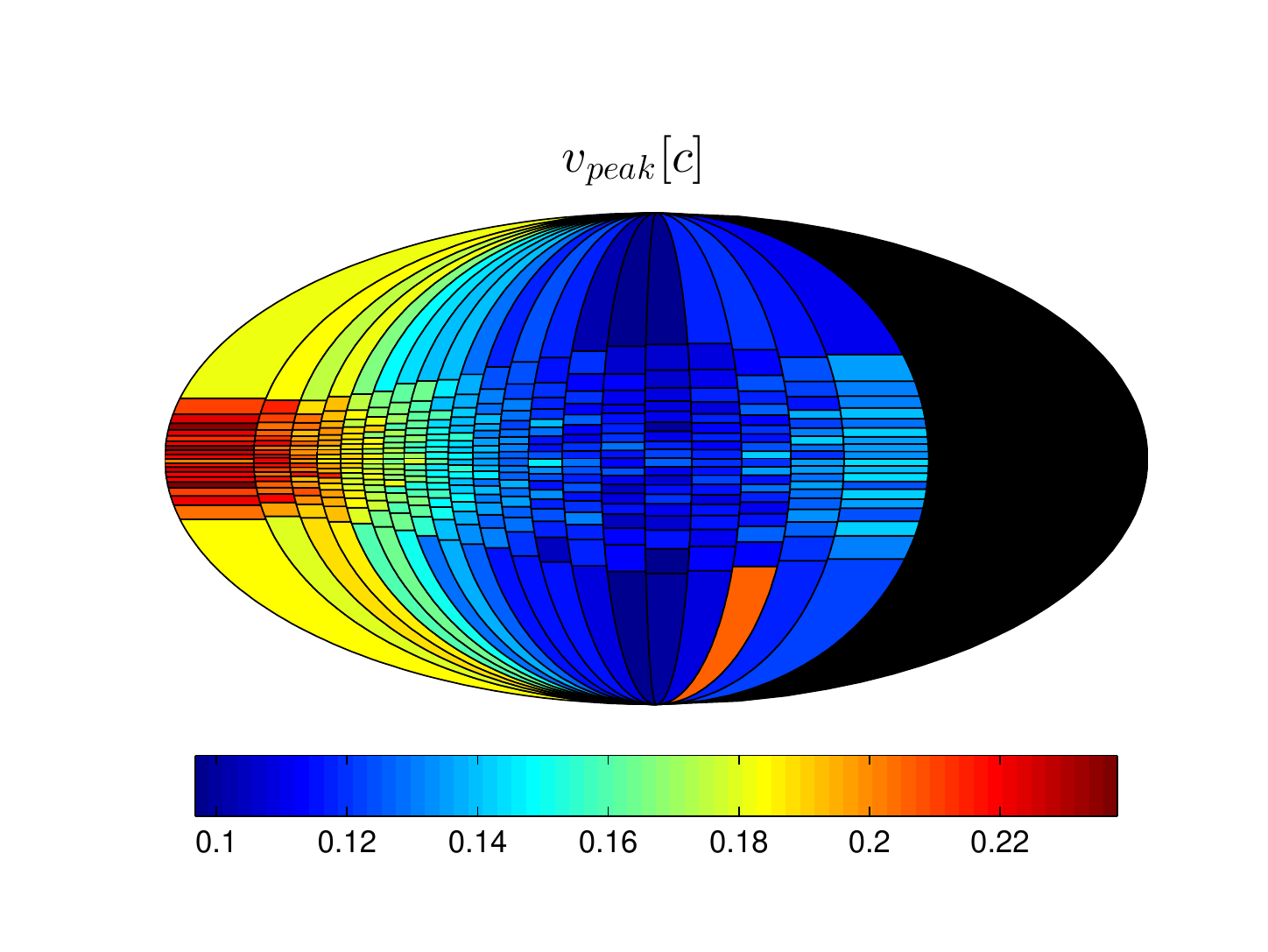,angle=0,width=0.45\textwidth} }
\end{subfigure}
~
\begin{subfigure}[light-curve comparison]{ \label{fig:GapRegion_lightcurve} \epsfig{file=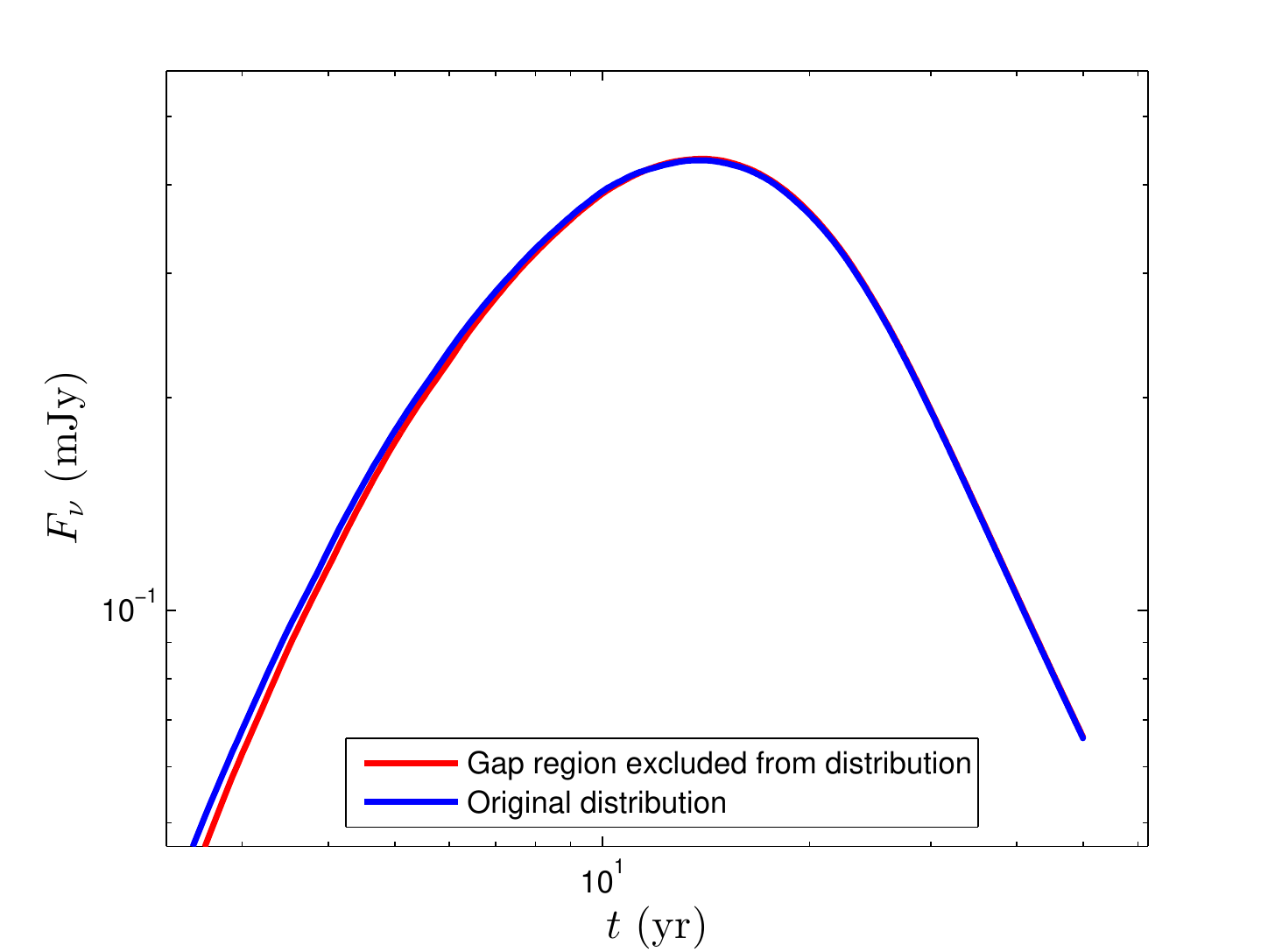,angle=0,width=0.45\textwidth} }
\end{subfigure}
\caption{(a) A velocity map of the ns16ns12 system, excluding the gap region. The black area corresponds the excluded region. This distribution is rotated with respect to Fig.\ \ref{fig:ns16ns12_Mollweide_velocity} (see Fig.\ \ref{fig:ns16ns12_equatorial_distribution_gap}). The figure illustrates that the velocity varies smoothly and monotonically from right to left apart from the jump across the excluded black region. (b) A comparison of the ns16ns12 light-curves produced by the original distribution and by the distribution excluding the gap region. The two light-curves are nearly identical, indicating that the gap region does not affect the results.}
\end{figure}

We show that the gap region in the equatorial distribution does not affect the resulting light-curve significantly, and therefore does not introduce any binning bias. This is shown by excluding the gap region from the analysis and comparing the results with and without this region. Figure \ref{fig:ns16ns12_equatorial_distribution_gap} illustrates the SPH particle distribution in the equatorial plane for the renewed analysis. The shaded region corresponds to the excluded angle encompassing the gap in the distribution, and the radially extending grey curves show the renewed binning of the system. The system is rotated clockwise by $1.95 ~\text{rad}$ with respect to the original distribution (depicted in Fig.\ \ref{fig:ns16ns12_equatorial_distribution}) so that the excluded region ends at $\varphi=2\pi$ and the first bin begins at $\varphi=0$. Figure \ref{fig:GapRegion_mollweide} continues by plotting the characteristic velocity map of this system, where the black region corresponds to the gap which has been excluded from the analysis. The velocity in this case varies smoothly and monotonically from right to left as the bins are dominated by faster moving ejecta. Finally,  Fig.\ \ref{fig:GapRegion_lightcurve} depicts the light-curves  for the original distribution, and the distribution from which the gap region has been excluded. It is evident from this figure that the gap region is inconsequential to the results, and that including it in the original analysis does not affect or bias the results. This is clear as the two light-curves presented in Fig.\ \ref{fig:GapRegion_lightcurve} overlap such that they are nearly indistinguishable. Additionally, we show more generally in Appendix \ref{subsec:Errors} that the arbitrariness of any binning choice does not affect the results for any of the three merger scenarios.

\begin{figure*}[h]
\centering
\begin{subfigure}[ns16ns12]{ \label{fig:lightcurves_a} \epsfig{file=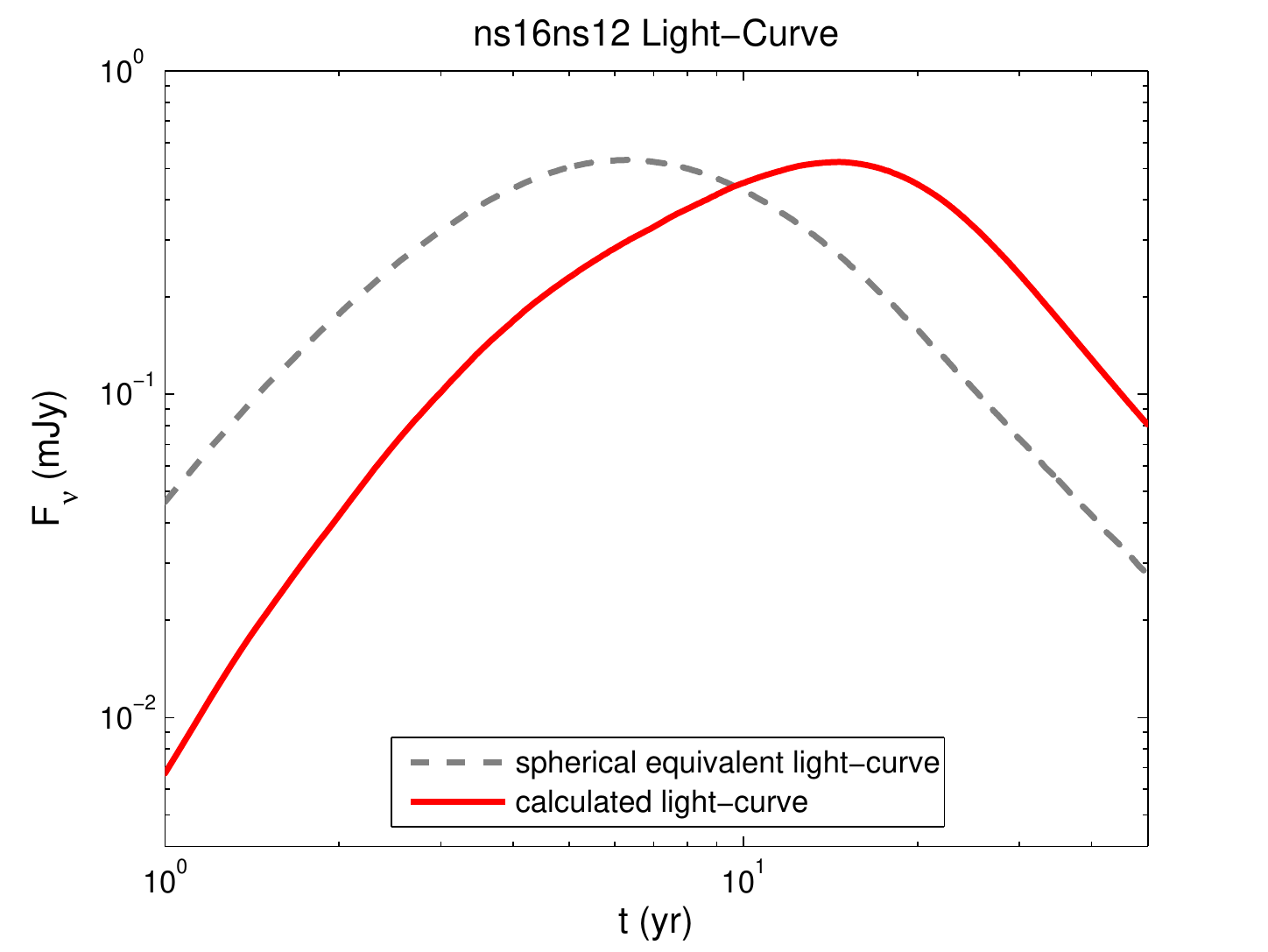,angle=0,width=0.45\textwidth} }
\end{subfigure}
~
\begin{subfigure}[ns14ns13]{ \label{fig:lightcurves_b} \epsfig{file=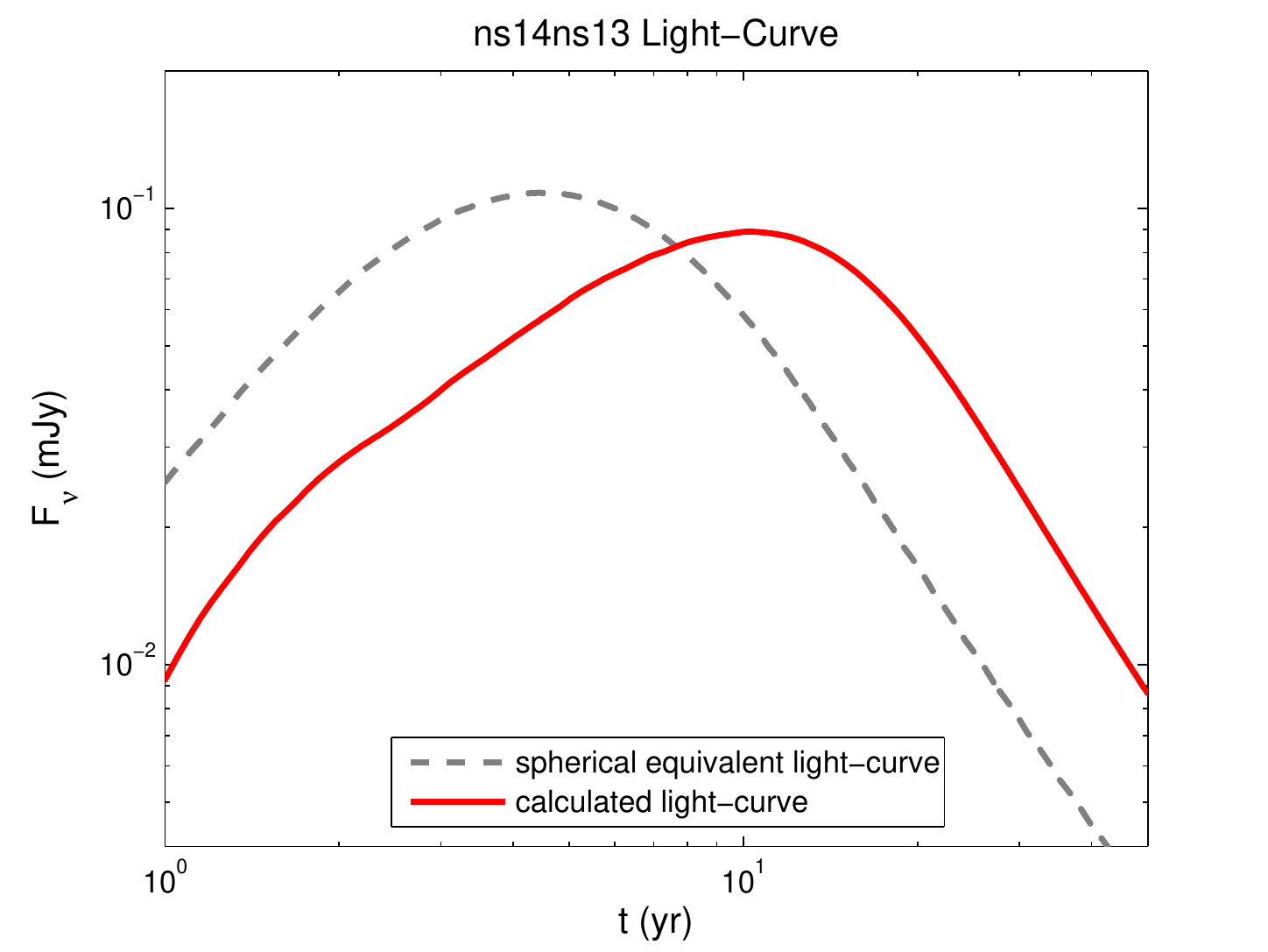,angle=0,width=0.45\textwidth} }
\end{subfigure}
~
\begin{subfigure}[ns14ns14]{ \label{fig:lightcurves_c} \epsfig{file=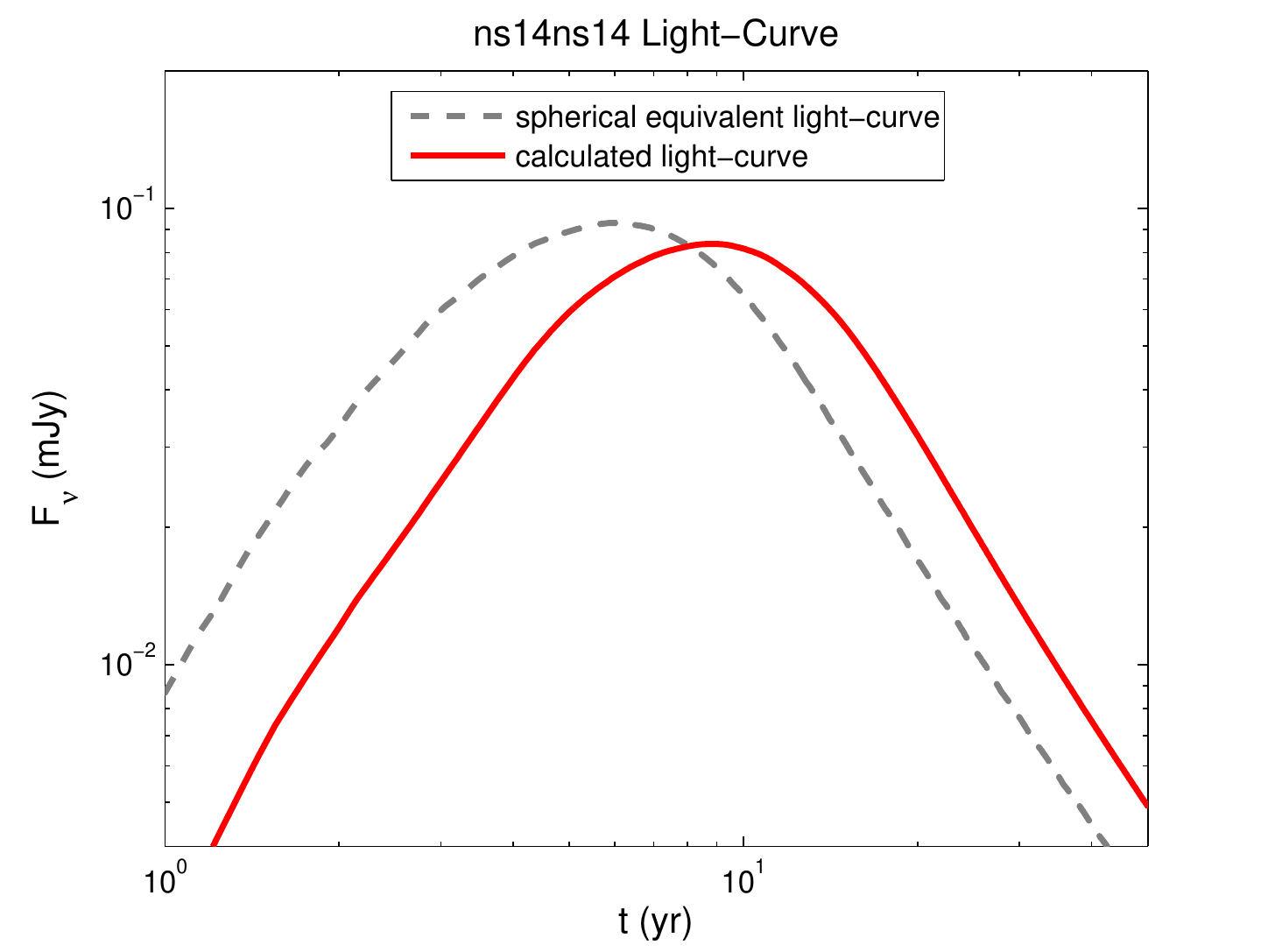,angle=0,width=0.45\textwidth} }
\end{subfigure}
\caption{The light-curves for the three binary merger systems. The flux is calculated at $\nu_{obs}=1.4 ~ \text{GHz}$ (which satisfies the relation $\nu_m,\nu_a<\nu_{obs}$ for all bin components) and taking a typical distance of $d=10^{27} ~ \text{cm}$. The dashed grey lines correspond to the spherical equivalent light-curves for these systems, which we compare to calculations of the present work (red lines). In all three cases the light-curves peak at later times than their spherical equivalents, and attain roughly the same peak fluxes. These two effects increase in their significance for larger binary mass ratios, which may be explained since these produce less symmetric outflows. Both these results are well explained theoretically (see \S\ \ref{sec:Analytic_Estimates}) if the characteristic velocities do not vary significantly between solid angle bins (see Fig.\ \ref{fig:Mollweide} and \ref{fig:TimeScale_Correlation}). Quantitative results for the peak time and flux are presented in table \ref{tab:ResultsTable}, along with these signals detectability prospects (see \S\ \ref{sec:Detectability}).} \label{fig:lightcurves}
\end{figure*}

\subsection{The Resulting Light-Curves} \label{subsec:Numerical_Results_subB}
After examining the angular binning and the distribution of characteristic variables we calculate the dynamic evolution of each angular bin according to it's isotropic equivalent evolution, and 
the resulting synchrotron light-curves. The light curves are compared to their spherical equivalents in Fig.\ \ref{fig:lightcurves}. It is not surprising that the less symmetric the underlying system is, the greater is the differences between the results and their spherical equivalents. In particular the peak time increases dramatically in the ns16ns12 system when compared with the timescale increase of the ns14ns14 merger. The basic reason for this result is the fact that nearly all of the ns16ns12 outflow is confined to the binary plane, and moreover it is significantly unevenly distributed within this plane, as most of the energy is concentrated in a small tidally disrupted tail. Thus, most of the system's energy is distributed over a small fraction of the sky, leading to a significantly larger energy density $dE/d\Omega$ when compared with the spherical equivalent situation (in which the energy is distributed evenly over $4\pi ~\text{rad}^2$). As the peak time is proportional to $(dE/d\Omega)^{1/3}$, this immediately translates into an increase in the light-curve's timescale. This reasoning works only as long as the characteristic velocity does not vary significantly, so that the timescale is determined almost entirely by the energy density.

Indeed, when looking back at figures \ref{fig:ns16ns12_Mollweide} and \ref{fig:ns16ns12_Mollweide_velocity} we notice that while the energy density fluctuates over three orders of magnitude, the velocity varies by a modest factor of two. Thus, to a zeroth order, the main effect contributing to the timescale delay is the compression of the ejected matter into a narrow solid angle which gives rise to a dramatic increase in the energy density $dE/d\Omega$ (or equivalently to the mass density $dM/d\Omega$).

\begin{figure}[]
\centering
\begin{subfigure}[]{ \label{fig:TimeScale_Correlation_a} \epsfig{file=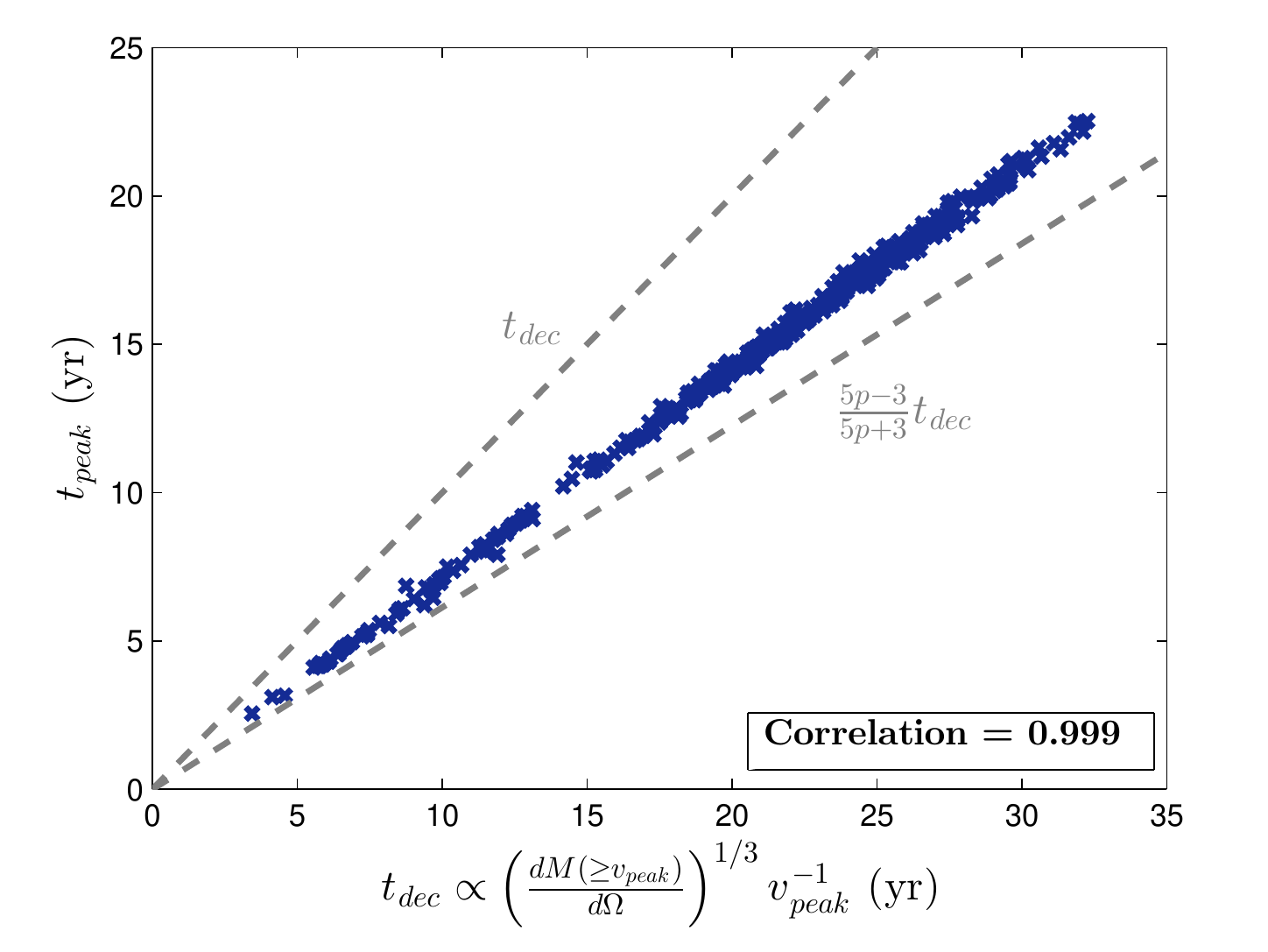,angle=0,width=0.45\textwidth} }
\end{subfigure}
~
\begin{subfigure}[]{ \label{fig:TimeScale_Correlation_b} \epsfig{file=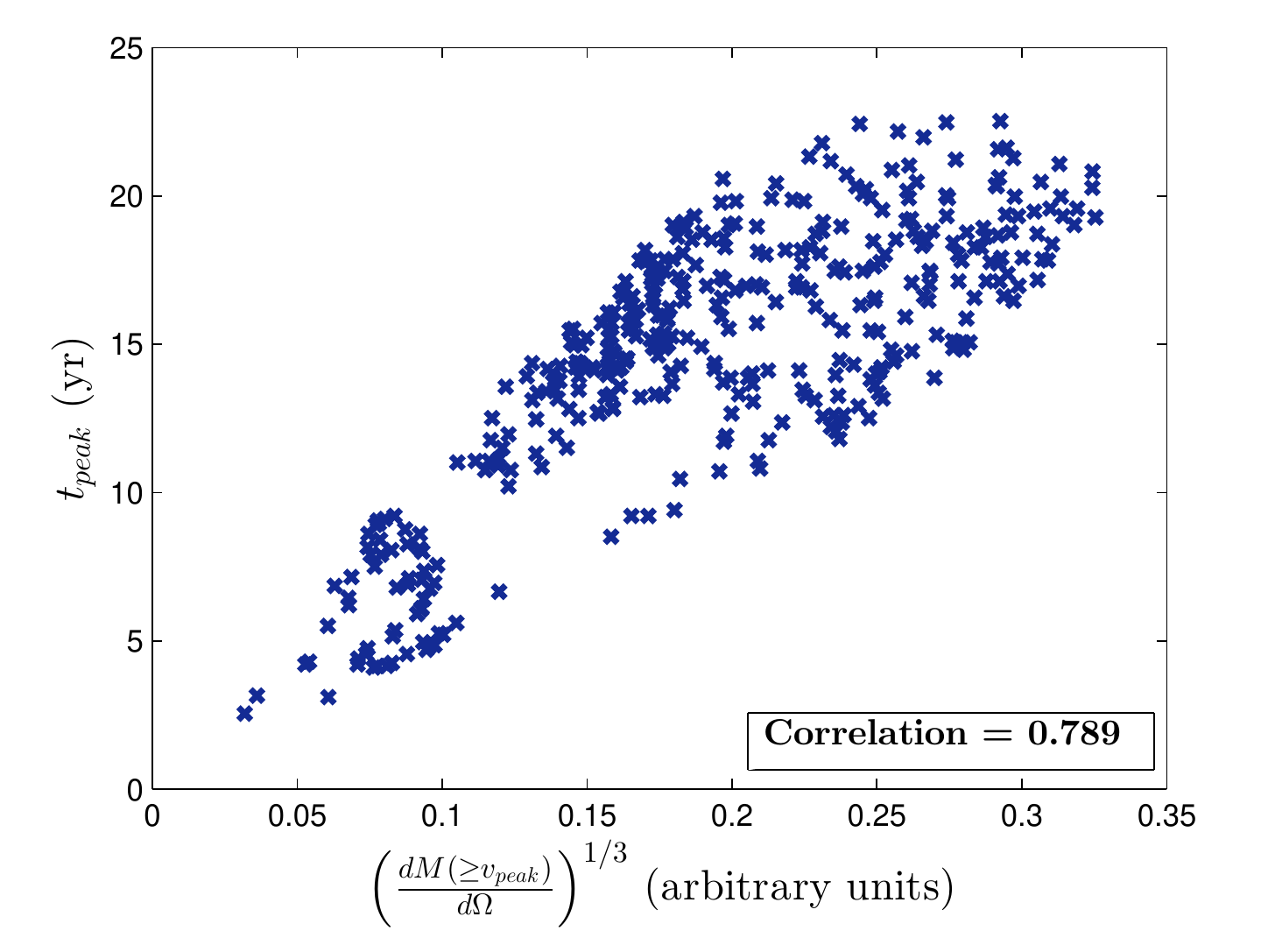,angle=0,width=0.45\textwidth} }
\end{subfigure}
~
\begin{subfigure}[]{ \label{fig:TimeScale_Correlation_c} \epsfig{file=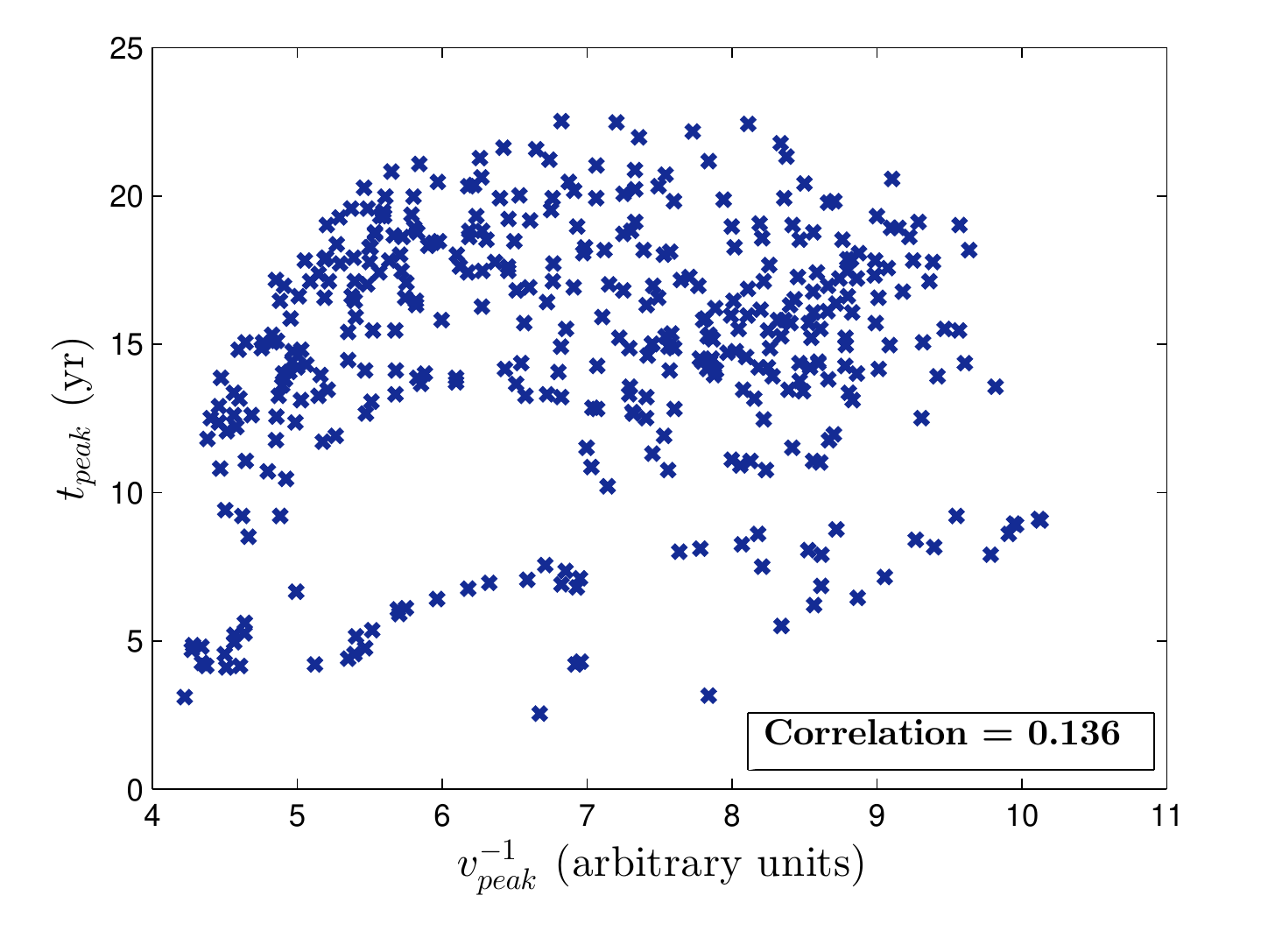,angle=0,width=0.45\textwidth} }
\end{subfigure}
\caption{The correlation of each solid angle bin's peak time with its mass and velocity distributions for the ns16ns12 system. Fig.\ \ref{fig:TimeScale_Correlation_a} clearly shows that the peak time scales as $t_{dec}$. Figures \ref{fig:TimeScale_Correlation_b} and \ref{fig:TimeScale_Correlation_c} illustrate that the main factor determining the timescale for this system is the mass density rather than velocity. This is quantified via the Pearson Correlation between the variables in each case. The fact the peak time is determined mainly by the mass density allows a comparison with the theoretical expectations in case the specific energy is fixed (see \S\ \ref{sec:Analytic_Estimates}). The dashed grey lines in Fig.\ \ref{fig:TimeScale_Correlation_a} show the upper and lower bounds on $t_{peak}$, as given by Eq.\ \ref{eq:peakTime_upperbound} and \ref{eq:peakTime_lowerbound}. The points are all well within these limits with very little scatter, indicating that the functional form of the cumlative energy distribution is nearly identical for all bins.} \label{fig:TimeScale_Correlation}
\end{figure}

In order to verify that the timescale is mainly influenced by the  concentration of the outflow into smaller solid angles, we plot in Fig.\ \ref{fig:TimeScale_Correlation} the peak time of each solid angle's signal versus the two independent variables contributing to the peak time, $dM(\geq v_{peak})/d\Omega$ and $v_{peak}$. Here, we choose to parameterize the peak time using the mass instead of energy density, since the energy density correlates by definition with the characteristic velocity, whereas the mass should be independent of it. From Eq.\ \ref{eq:peakTime_upperbound} and \ref{eq:peakTime_lowerbound} we find that the peak time scales with the deceleration time, i.e. $t_{peak} \propto \left( dM(\geq v_{peak}) / d\Omega \right)^{1/3} v_{peak}^{-1}$. Figure \ref{fig:TimeScale_Correlation_a} shows clearly that this relationship holds quite strongly, where the small deviations arise from the fact that an additional degree of freedom (which depends on the detailed energy distribution $E(\geq v_{peak})$) determines the exact value of $t_{peak}$. In any case it is evident from this figure that $t_{peak}$ lies within the lower and upper bounds given by Eq.\ \ref{eq:peakTime_lowerbound} and \ref{eq:peakTime_upperbound} which are illustrated as dashed grey lines in Fig.\ \ref{fig:TimeScale_Correlation_a}.

Figures \ref{fig:TimeScale_Correlation_b} and \ref{fig:TimeScale_Correlation_c} show the dependence of the peak time  on each one of the parameters independently. Also shown is  the Pearson Correlation between the two variables. Clearly, $t_{peak}$ depends strongly on the mass density and only very weakly on the velocity distribution. This supports our earlier conclusion that the timescale is governed by the ejecta's concentration into smaller portions of the sky, and not by dramatic changes in the characteristic velocities.

Finally, we compare the numerical results with the results of \S\ \ref{sec:Analytic_Estimates}, in which we have shown  that the flux weighted average of each solid angle's peak time must increase for any redistribution in which the specific energy is kept fixed. Figure \ref{fig:TimeScale_Correlation} shows that the peak time correlates only very weakly with the characteristic velocity, so that the specific energy is  inconsequential and may be treated as fixed. This justifies a comparison with the fixed specific energy scenario discussed in \S\ \ref{sec:Analytic_Estimates}. Still, in order to complete the comparison we must ascertain that the flux weighted mean peak time, $\tau$, is a valid  estimate of the total timescale. In \S\ \ref{sec:Analytic_Estimates} we argued that $\tau$ is a reasonable estimate as long as the different contributing light-curves overlapped significantly. This was quantified by estimating the necessary overlap width, $\Delta t$, for single-velocity shells. Non-single-velocity shells give rise to larger light-curve widths, yet we can still use the single-velocity shell results as lower bounds on $\Delta t$ in this case.

\begin{figure*}
\centering
\begin{subfigure}[ns16ns12]{ \label{fig:MeanTime_ns16ns12} \epsfig{file=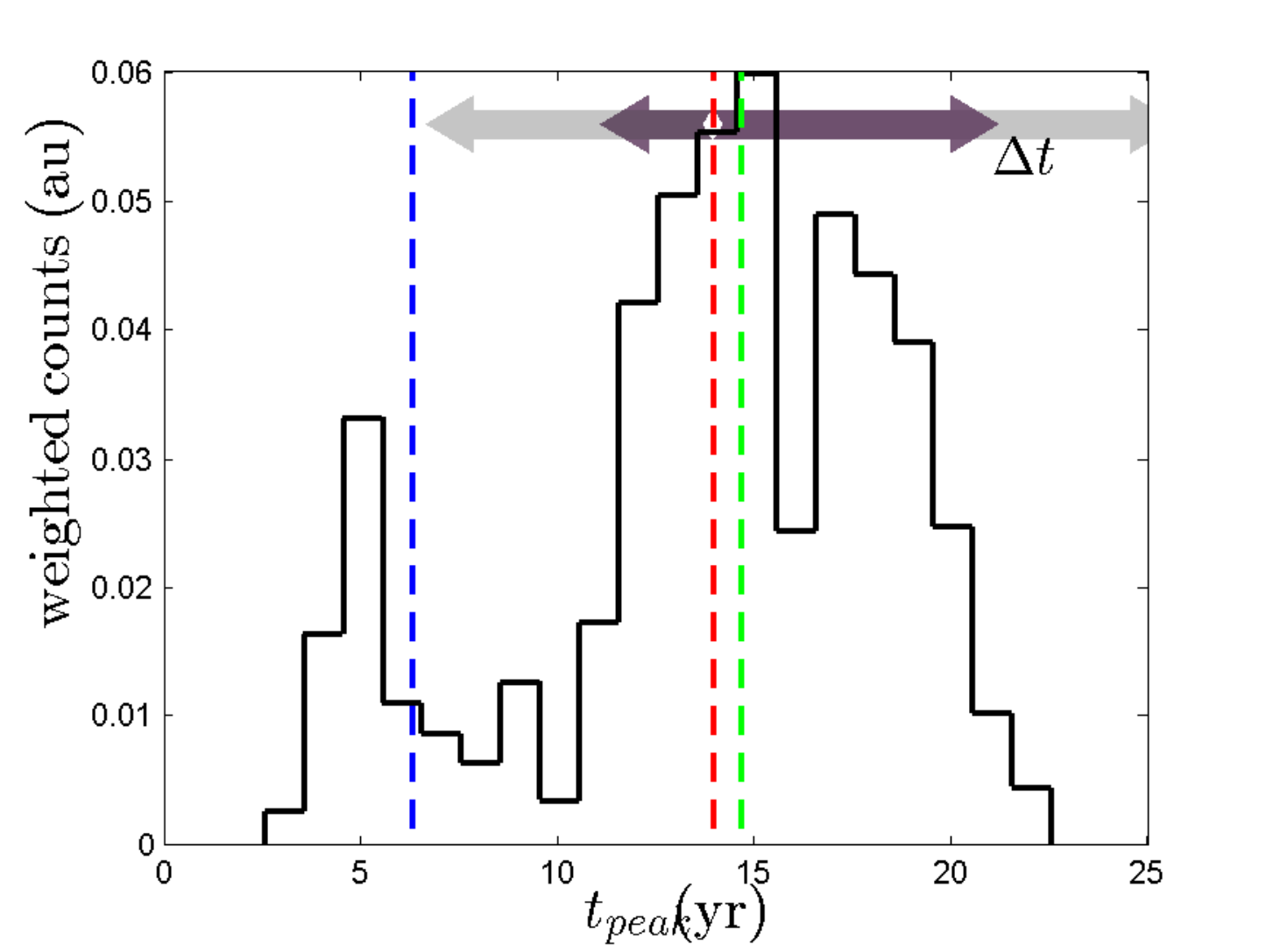,angle=0,width=0.45\textwidth} }
\end{subfigure}
~
\begin{subfigure}[ns14ns13]{ \label{fig:MeanTime_ns14ns13} \epsfig{file=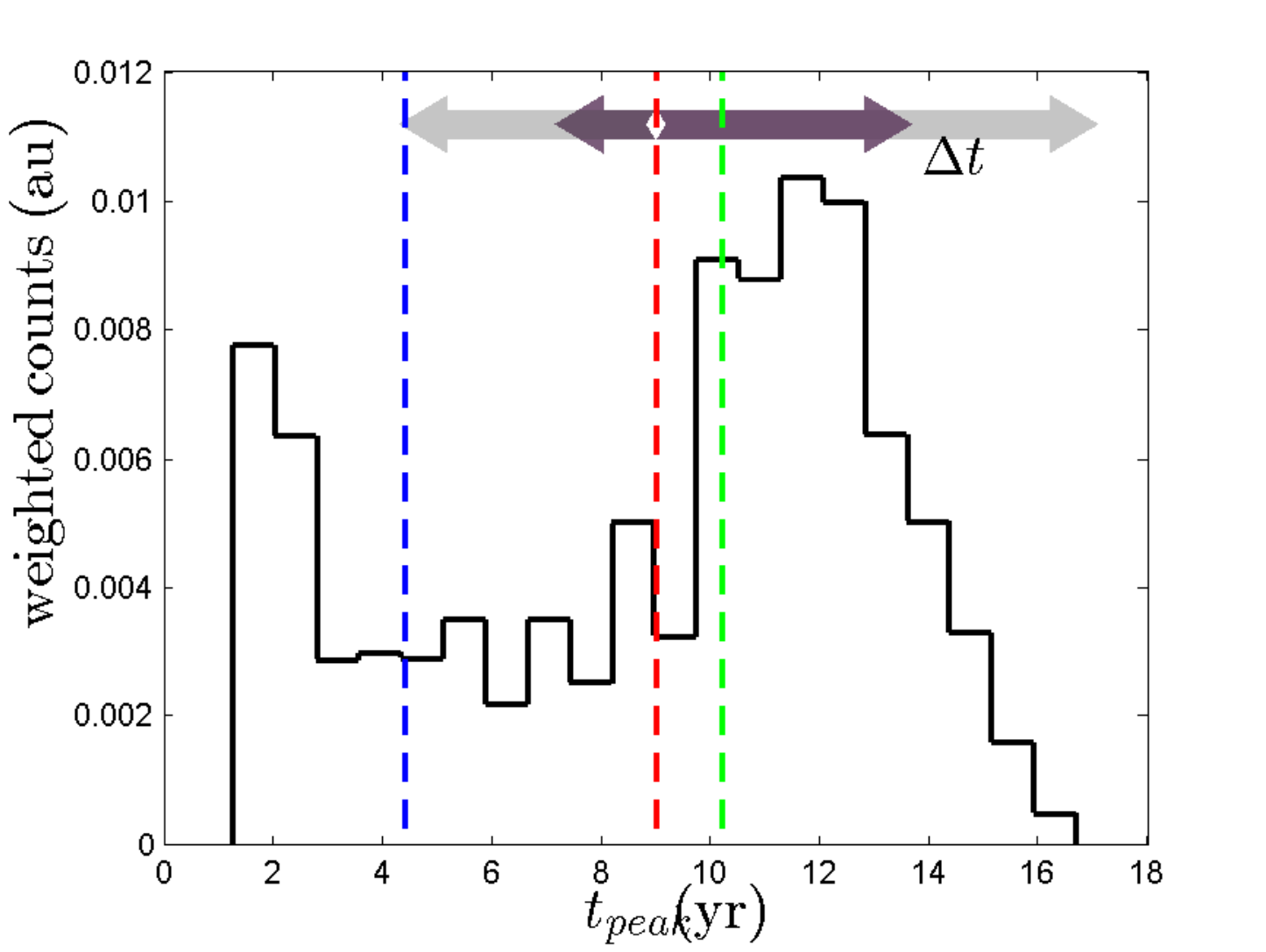,angle=0,width=0.45\textwidth} }
\end{subfigure}
~
\begin{subfigure}[ns14ns14]{ \label{fig:MeanTime_ns14ns14} \epsfig{file=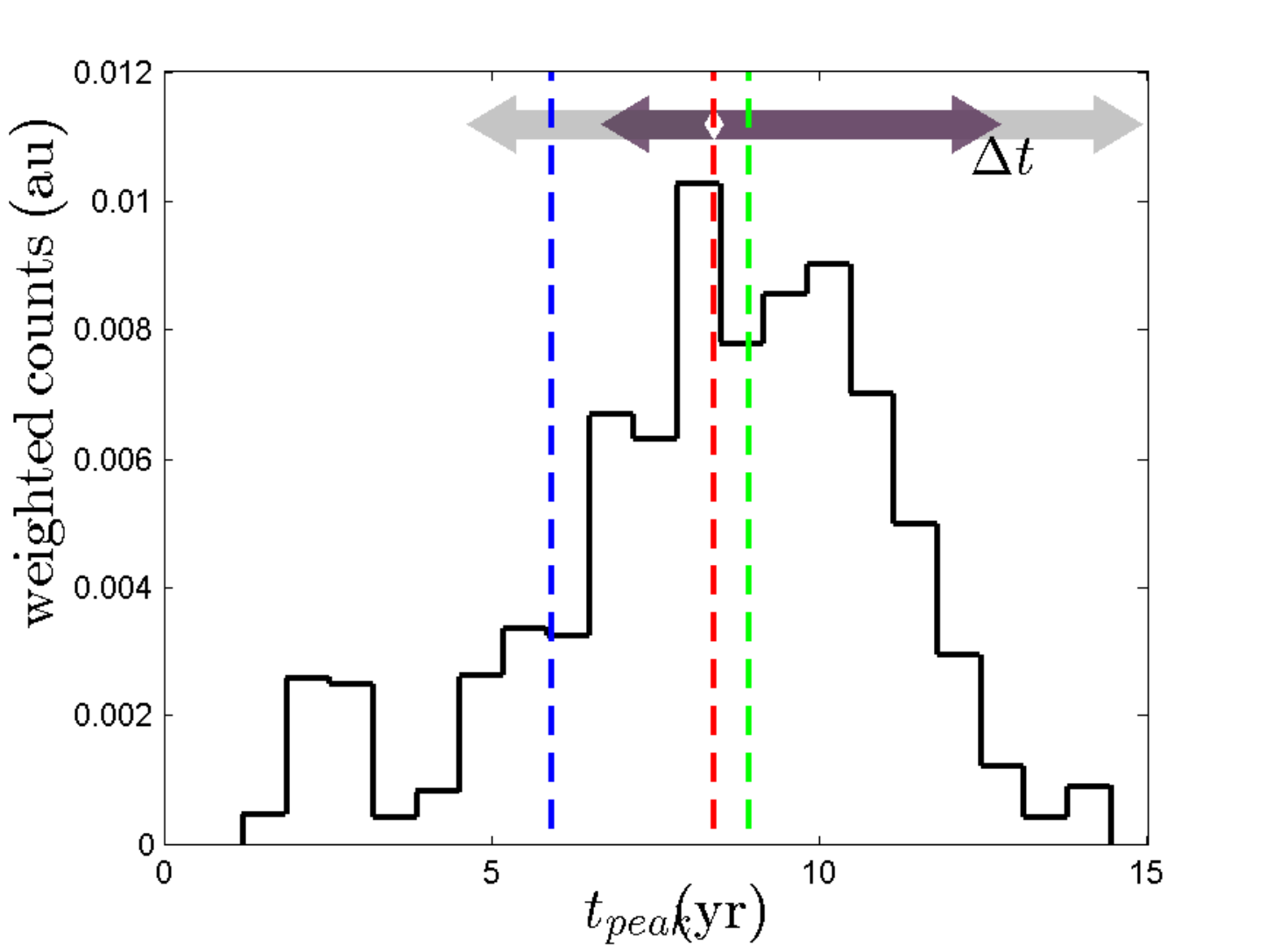,angle=0,width=0.45\textwidth} }
\end{subfigure}
~
\begin{subfigure}{ \epsfig{file=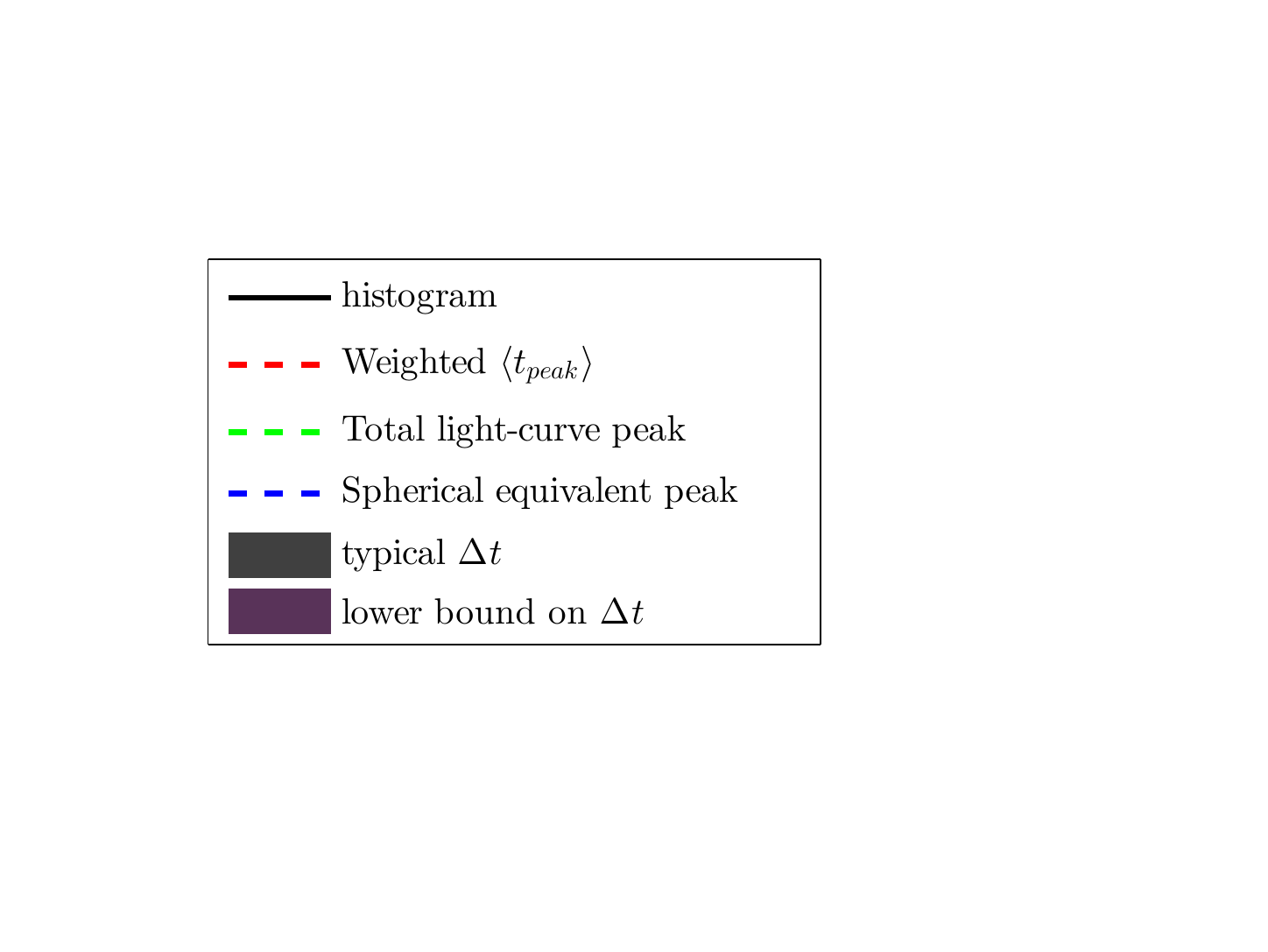,angle=0,width=0.45\textwidth} }
\end{subfigure}
\caption{The flux weighted distributions of peak times for each solid angle component for  the three  systems. The red dashed lines indicate the weighted mean of these components. The green dashed lines depict the actual peak time of the total light-curve. The blue dashed lines show the peak time in the spherical equivalent case. Arrows depict the light-curves' width around the weighted mean, defined as the times at which the flux diminishes to half it's peak value. Dark/short arrows show the lower bounds on this width (Eq.\ \ref{eq:t-} and \ref{eq:t+}), whereas light/long arrows show the typical widths for our systems. The figures show that the weighted mean peak time is a good estimate of the total light-curve's peak time. This is consistent with the fact that the underlying components overlap significantly (their width's overlap over some region).} \label{fig:MeanTime}
\end{figure*}

Figure \ref{fig:MeanTime} depicts the flux weighted histograms of each bin's peak time for the three  systems , as well as the additional relevant timescales of the problem. Also illustrated are arrows showing the width, $\Delta t$, of of the underlying light-curves around the weighted mean. Short/dark arrows show the lower bounds on this width (as given by Eq.\ \ref{eq:t-} and \ref{eq:t+} for single-velocity shells), and the longer/lighter arrows show the actual measured width of a typical component's light-curve. The distribution of widths is centered around $t_{1/2}^{(-)} \sim 0.5 \times t_{peak}$, and $t_{1/2}^{(+)} \sim 1.8 \times t_{peak}$. It is clear that most of the components significantly overlap and hence we should expect the weighted mean to serve as a reasonable estimate of the total timescale. This is also apparent directly by comparing the calculated weighted mean (depicted as a dashed red line in these figures) with the actual total light-curve peak time (depicted as dashed green lines). In all three cases these curves are in proximity of each other when compared with the spherical equivalent peak time (dashed blue line). Another point worth mentioning is the fact that the actual peak time is always greater than the weighted mean estimate. This is most likely caused by the asymmetry of the light-curve  around the peak  (the light-curve varies at late/early times as $t^{-1.65}$/$t^3$ respectively). As this asymmetry is unaccounted for by the weighted mean peak time, it can be expected that the actual timescale be delayed slightly further than this estimate would suggest.

Analyzing `realistic' merger dynamical outflows using our piecewise spherical approximation, we have  found an increase in the timescale of the merger radio light-curves in comparison with their spherical equivalent estimates 
studies in previous works \citep{NakarPiran2011,Piran2013}. This shifts the peak timescales to the order of $\sim 10$ years. In contrast, the peak flux of the light-curves is comparable to it's spherical equivalent and it does not change significantly when taking into account the nontrivial outflow geometry. Both these results are consistent with the analytic estimates presented in \S\ \ref{sec:Analytic_Estimates} provided that the characteristic velocities of various solid angles are treated as roughly constant. Numerical values for the peak flux and peak timescales found  are given in table \ref{tab:ResultsTable}, which also states the characteristic velocities and energies for the spherical equivalent cases.

\section{Detectability} \label{sec:Detectability}
We turn now to discuss the detectability of compact binary mergers' radio flares based on these results.
The number of radio flares observable by an instrument with detection level $F_{lim}$ in an all sky snapshot is
\begin{equation} \label{eq:N_simple}
N_{all-sky} \approx \mathcal{R} V \Delta t ,
\end{equation}
where $\mathcal{R}$ is the merger event rate, $V$ is the detectable volume and $\Delta t$ is the time that the flux is above the detection threshold $F_{lim}$. We can improve on Eq.\ \ref{eq:N_simple} by taking into account the dependence of $\Delta t$ on the distance to the merger event as follows:
\begin{align} \label{eq:N_integral}
&N_{all-sky} = \int_0^\infty \Theta\left( F\left(r\right) - F_{lim} \right) \mathcal{R} \Delta t\left(r\right) 4\pi r^2 dr = \\ \nonumber &= \mathcal{R} \frac{4\pi}{3} \left(\frac{L}{4\pi F_{lim}}\right)^{3/2} \int_1^\infty \frac{3}{2} \Delta t\left( \chi \right) \chi^{-5/2} d\chi \equiv \mathcal{R} V \left\langle \Delta t \right\rangle .
\end{align}
Here $L$ is the merger's  peak luminosity, $\chi \equiv F(r)/F_{lim}$, and we assumed isotropy and neglected any cosmological corrections to the merger rate $\mathcal{R}\left(r\right)$ and peak flux $F(r)=L/(4\pi r^2)$. $\Delta t\left(\chi\right)$ is the time the flux is above the detection limit, and it depends on the exact shape of the radio flare light-curve as well as on the ratio between the peak flux $F$ and the detection threshold $F_{lim}$. In the simple case of a single-velocity shell we can write down $\Delta t\left(\chi\right)$ explicitly using the light-curve scaling relations given by Eq.\ \ref{eq:SingleVelocity_Flux} (assuming again that $\nu_m,\nu_a<\nu_{obs}$),
\begin{equation}
\Delta t\left(\chi\right) = \left( \chi^{\frac{10}{15p-21}} - \chi^{-1/3} \right)t_{dec} \ .
\end{equation}
Using this result we continue in finding the relevant time-scale $\left\langle \Delta t \right\rangle$:
\begin{equation} \label{eq:delta_t_singlevelocity}
\left\langle \Delta t \right\rangle = \frac{18(5p+3)}{11(45p-83)} t_{dec} \approx 0.86 ~t_{dec}
\end{equation}

Plugging this expression into Eq.\ \ref{eq:N_integral} and simplifying the result gives an estimate for the number of merger remnants observable over the whole sky. This result is similar to that given by \cite{NakarPiran2011}, which estimated that $\Delta t \approx t_{dec}$, but is larger due to a round-off error in their calculations. Expressing $t_{dec}$ and $L$ in terms of the single-velocity outflow energy $E$ and velocity $\beta c$ using Eq.\ \ref{eq:F_m} and \ref{eq:t_dec}, we find that the the number of spherically symmetric single-velocity shell radio remnants observable at $1.4 \text{GHz}$ is:
\begin{align} \label{eq:N_single_veloctiy}
N_{all-sky}^{1.4 \text{GHz}} \approx 123 ~&\left(\frac{E}{10^{50} \text{erg}}\right)^{11/6} \left(\frac{\epsilon_B}{0.1}\right)^{\frac{3(p+1)}{8}} \left(\frac{\epsilon_e}{0.1}\right)^{\frac{3(p-1)}{2}} \nonumber \\
&\times n^{\frac{9p+1}{24}} \left(\frac{\beta}{0.2}\right)^{\frac{45p-83}{12}} \left(\frac{F_{lim}}{0.1 \text{mJy}}\right)^{-3/2} \nonumber \\ 
&\times \left(\frac{\mathcal{R}}{300 \text{Gpc}^{-3} \text{yr}^{-1}}\right) \left(\frac{\left\langle \Delta t \right\rangle / t_{peak}}{0.86}\right)
\end{align}

We use our own canonical values in evaluating this expression, which are somewhat different than those chosen by PNR13. We notice that the expression depends sensitively on the outflow velocity: $N_{all-sky} \propto \beta^{2.46}$ which we take to  be slightly lower than $0.2c$. $N_{all-sky}$ depends also on the ejecta energy which should be in the range of $E \sim 10^{50}-10^{51} \text{erg}$ for the outflows considered. A larger value than our canonical $10^{50} \text{erg}$ will  cause an increase in $N_{all-sky}$ compared with our canonical value. Additionally, these results are sensitive to the ISM density ($N_{all-sky} \propto n^{0.98}$) so that binaries merging in less tenuous surroundings will decrease the detectability as well. Based on  current observation of Galactic neutron star binaries we expect that at least a significant fraction of them be located in the disk of their host galaxy, where the typical ISM densities are $\sim 1 \text{cm}^{-3}$ \citep{Draine2011}. A caveat to this point is that it stems from the observed Galactic binary neutron star population, which is both small, and confined within Milky-Way alone, so that other values of $n$ are quite possible and this parameter is left ill-constrained. For example, \cite{Fong2014} find $n \approx 5 \times 10^{-3}-30$ by evaluating the 130603B short GRB afterglow. This case illustrates the large uncertainty in ascertaining the circum-binary density using sGRB afterglow observations, and although kicks could relocate neutron star binaries outside their host galaxy where $n$ would be quite small, there is no clear evidence to rule out the galactic merger scenario.

Using these light-curves   
we can estimate the detectability for each merger case by evaluating Eq.\ \ref{eq:N_integral} numerically. The results are given in table \ref{tab:ResultsTable}, which also states the peak flux and peak time found for the different merger light-curves. Additionally the table includes the values of $\left\langle \Delta t \right\rangle / t_{peak}$ which are the corrections to the approximation $\Delta t \approx t_{peak}$ taking into account the specific light-curve shape. These calculated values (which are always of order unity) should be taken with a grain of salt because these light-curves may not necessarily reproduce the exact light-curves due to approximations   used (which may break down after the ejecta begins it's deceleration at $t_{peak}$). In any case, a conservative lower bound on $N_{all-sky}$ is given by replacing $\left\langle \Delta t \right\rangle / t_{peak}$ with the single velocity shell value of $0.86$ (see Eq.\ \ref{eq:delta_t_singlevelocity}). This substitution is straightforward since $N_{all-sky}$ scales linearly with $\left\langle \Delta t \right\rangle / t_{peak}$.

We also provide in table \ref{tab:ResultsTable} the outflow velocity at $t_{peak}$ and the energy of the outflow above this velocity for the spherical equivalent cases. As explained in \S\ \ref{sec:Spherical_Systems}, these are the characteristic energy and velocity which allow us to estimate the peak flux and time using single-velocity shell results (by exchanging $E,v$ with $E\left(\geq v_{peak}\right), v_{peak}$ in the relevant equations). Substituting these characteristic values into Eq.\ \ref{eq:N_single_veloctiy} for instance, yields similar results to $N_{all-sky}$ calculated numerically. The only reason the results do not coincide exactly is the uncertainty in evaluating the peak time using the characteristic velocity and energy. This time can be found using these two variables only up to a factor of order unity (and confined between $0.6-1$, see Eq.\ \ref{eq:peakTime_upperbound}, \ref{eq:peakTime_lowerbound}).

\begin{table*}
\begin{tabular}{c || c  c  c  c  c  c}
 & $F_{\nu}^{peak} [\text{mJy}]$ & $t_{peak} [\text{yr}]$ & $N_{all-sky}$ & $\left\langle \Delta t \right\rangle / t_{peak}$ & $v_{peak} [c]$ & $E\left(\geq v_{peak} \right) [\text{erg}]$ \\ \hline
 ns16ns12 & $0.53$ & $14$ & $9.0 \times 10^3$ & $1.22$ & & \\
 Spherical Equivalent & $0.53$ & $6.3$ & $5.2 \times 10^3$ & $1.59$ & $0.187$ & $7.04 \times 10^{50}$ \\ \hline
 ns14ns13 & $0.090$ & $10.4$ & $563$ & $1.49$ & & \\
 Spherical Equivalent & $0.108$ & $4.4$ & $281$ & $1.32$ & $0.174$ & $1.75 \times 10^{50}$ \\ \hline
 ns14ns14 & $0.084$ & $8.7$ & $391$ & $1.38$ & & \\
 Spherical Equivalent & $0.093$ & $5.9$ & $287$ & $1.27$ & $0.153$ & $2.13 \times 10^{50}$
\end{tabular}
\caption{Numerical results illustrating the detectability of the  systems considered. Each row designates one of the three  systems and is divided into a top portion stating the results of the current analysis, and bottom portion stating the results for the spherical equivalent systems. The results are calculated at an observed frequency of $\nu_{obs}=1.4 \text{GHz}$, assuming a source distance of $d=10^{27} \text{cm}$, ISM density of $n=1\text{cm}^{-3}$, and microphysical parameters $\epsilon_e=\epsilon_B=0.1$ and $p=2.5$. The detectability is estimated by the number of observable sources in an all sky snapshot, $N_{all-sky}$, assuming a merger rate of $\mathcal{R}=300 \text{Gpc}^{-3} \text{yr}^{-1}$ and a detection threshold of $F_{lim}=0.1 \text{mJy}$. For the spherical equivalent systems only, the characteristic velocity and energy are stated in the rightmost columns.} \label{tab:ResultsTable}
\end{table*}

Our results for $N_{all-sky}$ are significantly larger than those estimated by previous works \citep{NakarPiran2011,Piran2013}, even for the spherical equivalent cases. There are several reasons for this increase. The first is a calculation round-off error in these papers, which leads to a factor of $\sim 5$ difference in Eq.\ \ref{eq:N_single_veloctiy} ($\sim 20$ as estimated in \cite{NakarPiran2011} as opposed to $\sim 110$ given in this work, taking $\left\langle \Delta t \right\rangle = t_{peak}$ and the canonical variables of \cite{NakarPiran2011} for the comparison). The second has to due with the fact that the typical energies and velocities of the merger outflows are different than those evaluated \cite{NakarPiran2011}. Specifically, for typical energies of $2\times 10^{50} \text{erg}$ and velocities of $0.2 c$ we get a further increase of $(2\times 10^{50}/10^{49})^{11/6}(0.2/1)^{2.46} \approx 4.6$ in the detectability. Lastly, for the case of the non-spherical light-curves calculated in this work, the peak time has increased by $\sim 2$ compared with the spherical equivalent cases, so that the number of mergers observed in an all sky snapshot  doubles.

A central issue in considering the detectability of any signal is it's identification. While $N_{all-sky}$ determines the number of signals above the detection threshold of a given radio facility, it is an entirely different question whether these signals can be identified as merger radio flares. The main difficulty our present results pose to the identification process is the increase in time-scale compared with previous works \citep{Piran2013}. The time-scales we estimate in this work are on the order of $\sim 10 ~\text{yr}$. This long period will severely influence the ability to identify these signals as transients and therefore associate them with merger outflows.

In order to identify the source signal as a transient, a significant change in it's flux must be measured. This is usually achieved by returning and pointing the radio telescope towards the source at several different times, but on the scale of a decade (on which the signal changes significantly), most instruments will have been upgraded or replaced by more advanced facilities, making any recurring measurements difficult to conduct and contaminated by instrumental changes. An alternative method of classifying a transient event is detecting smaller variations in the signal which would occur over smaller time periods. The problem with this method is that a typical detection should be very close to the detection threshold at a low S/N ratio, and would therefore not be sensitive enough to small variations in the flux. Thus, the only way such an approach could work were if an unlikely strong signal were detected, the chances of which are drastically diminished.

\subsection{Comparison with GRB Afterglows} \label{subsec:Comparison_with_GRB_afterglows}
We conclude this section by comparing the detectability of merger radio flares with the detectability of GRB orphan afterglows, which are often considered as more promising electromagnetic counterparts to mergers. The well constrained observables of GRBs are the isotropic equivalent energy of the burst emitted in gamma-rays $E_{\gamma , \mathrm{iso}}$, and the rate of observed GRBs $\mathcal{R}_{GRB}^{obs}$. Orphan afterglows on the other hand are produced after the beamed ultra-relativistic jet has expanded into a roughly spherical shell, emitting isotropic radiation. Afterglow luminosities and rates should therefore depend on the total energy of the GRB and overall GRB rate (not the observed rate which only include bursts pointing towards Earth). Assuming that the GRB emission mechanism is very efficient, the isotropic energy of the burst should be comparable to the isotropic energy emitted in gamma-rays, $E_\mathrm{iso} \sim E_{\gamma , \mathrm{iso}}$ \citep{Nakar2007}. The total burst energy and overall rate can then be calculated based on the well constrained $E_\mathrm{iso}$ and $\mathcal{R}_{GRB}^{obs}$, if we assume that the initial GRB jet covers a fraction $f_b$ of the sky:
\begin{equation} \label{eq:GRB_energy_rate}
E_{tot} = E_\mathrm{iso} f_b ~ ; ~~~ \mathcal{R}_{GRB} = \mathcal{R}_{GRB}^{obs} f_b^{-1} \ .
\end{equation}

We use these results in estimating the number of GRB orphan afterglows detectable at $1.4 ~\text{GHz}$, $N_{all-sky}^{orphan}$, similarly to the calculation for merger radio flares. The observed rate of GRBs is enery dependent, and can be approximated as a power-law over the energy range of $10^{49}-10^{51} \text{erg}$: $\mathcal{R}_{GRB}^{obs} \approx 10 ~\text{Gpc}^{-3} \text{yr}^{-1} \left(E_\mathrm{iso} / 10^{49} \text{erg}\right)^{-\alpha}$ where $\alpha \approx 0.5-1$ \citep{GuettaPiran2006,Nakar2006}. Plugging this luminosity function into Eq.\ \ref{eq:N_single_veloctiy}, and using Eq.\ \ref{eq:GRB_energy_rate} we find
\begin{align} \label{eq:GRB_N}
N_{all-sky}^{orphan} \approx 3.6 ~&f_b^{5/6} \left(\frac{E_\mathrm{iso}}{10^{49} \text{erg}}\right)^{11/6-\alpha} \left(\frac{\epsilon_B}{0.1}\right)^{\frac{3(p+1)}{8}}  \\ \nonumber
&\times \left(\frac{\epsilon_e}{0.1}\right)^{\frac{3(p-1)}{2}} n^{\frac{9p+1}{24}} \left(\frac{F_{lim}}{0.1 \text{mJy}}\right)^{-3/2} ,
\end{align}
where we take $\left\langle \Delta t \right\rangle / t_{peak} = 1$, and $\beta=1$ since the GRB outflow is ultra-relativistic.

Using an expression similar to Eq.\ \ref{eq:GRB_N}, \cite{Levinson2002} showed that although the GRB afterglow rate increases with smaller beaming factors, the overall detectability will decrease due to the decrease in the afterglow energy (see Eq.\ \ref{eq:GRB_energy_rate}). Most of the variables in Eq.\ \ref{eq:GRB_N} are well constrained, with the exception of the beaming factor $f_b$, which could be on the order of $f_b = 1/20$ (by definition $f_b<1$). Such small beaming factors would lower the number of observable orphan afterglows in an all sky snapshot by more than an order of magnitude, making a blind survey detection implausible.

Comparing this result with those calculated previously in \S\ \ref{sec:Detectability} we see that the detectability of GRB afterglows is two to three orders of magnitude lower than the detectability of merger radio flares. This result strengthens previous claims by PNR13 that radio flares produced by dynamically ejected material are more promising electromagnetic counterparts to compact binary mergers than GRB afterglows. One important caveat is the issue of differing timescales. While we have shown previously that the radio flare time-scale is of the order $\sim 10 ~\text{yr}$, GRB afterglows expand relativistically and will most likely be dominated by low energy bursts of $\sim 10^{49} ~\text{erg}$ (because of the observed luminosity function) so that they will peak on much shorter time-scales of $\sim 1 ~\text{week}$. This shorter time-scale will likely make the identification of GRB orphan afterglows in radio surveys significantly easier. Still, because of the large disparity between the detectability of the two signals we expect that merger radio flares will be easier to detect, especially in bind surveys.

\section{Discussion and Conclusions} \label{sec:Discussion_and_Conclusions}
We review here our main results and discuss their implications in a broader context. In particular we address the question of whether the radio flares found here show promise as detectable electromagnetic signals from compact binary mergers. We finish by suggesting further research directions into this and related problems.

Throughout this work we have studied radio signals resulting from the deceleration of merger dynamic outflows as they interact with the surrounding ISM. We have expanded on previous works \citep{NakarPiran2011,Piran2013} by taking into account the asymmetry in these outflows (which have previously been approximated as spherically symmetric). We focus on observed frequency $\nu_{obs}=1.4 ~\text{GHz}$ at which radio facilities are most sensitive, and at which the outflow is optically thin.

Although the treatment of the outflow dynamics are rather crudely approximated, we expect we have captured the `essence' of the effect in the analysis - namely that the asymmetric outflow confines the system's energy into a smaller portion of the sky, increasing the isotropic equivalent energy, which in turn increases the peak time.  Note, however, that our approximation maximizes the effect of deviation from spherical symmetry and as such these results are a sort of a limit on this effect.

Our analytic estimates show that asphericity can either increase or decrease the peak timescale, depending on the way mass and energy are distributed. A particularly important case is when specific energy remains constant. This situation describes aspherical outflows expanding with the same velocity in each direction, such that the asymmetry is caused by an uneven distribution of mass in different directions. More generally, this scenario reasonably approximates outflows in which the mass density, and not the velocity, is the dominant source of asymmetry. We therefore expect this to be the prevalent situation as long as the outflow does not exhibit a fast moving jet component, and in particular we find this describes well the ejecta found in merger simulations.
In this case the analytic estimates show an increase in the peak timescale with respect to the spherical equivalent light-curve. Additionally, the peak flux should remain roughly the same as in the spherical equivalent case. The physical argumentation for these results is rather simple - since  the outflow mass is confined into a small solid angle, it must expand further radially in order to accumulate a comparable amount of ambient mass, this takes longer (since the expansion velocity is unchanged). The synchrotron flux on the other hand is determined only by the amount (not distribution) of emitting mass and the velocity, which are both unchanged.

Applying our method to the ejecta distributions found by \cite{Rosswog2014}, we find that the light-curve's peak time increases by a factor of $\sim 2$ compared with the spherical approximation, and is typically on the order of $\sim 10$ years, for an ISM density of $n=1 \text{cm}^{-3}$. The ISM density is an ill constrained parameter and could be orders of magnitude lower if the merger occurs outside it's host galaxy disk. In such cases the timescale will be even longer than estimated above, as $t_{peak} \propto n^{-1/3}$. We additionally find that the light-curve's peak flux remains roughly the same as in the approximated spherical treatment.

The implications of our findings are important in assessing the detectability prospects of these signals. The longer timescale is a double edged sword in addressing this issue. On one hand, longer timescales  make the identification  more difficult. 
This is important for both all sky searches and for attempts to identify 
radio flares following identification of a GW candidate.
On the other hand, the longer timescales mean a larger number of observable sources in an all sky snapshot. The combined effect will depend on the telescope used and on the observation strategy as well as the unknown background field of radio transients that might compete with these events.

Support for this work was provided by an ERC advanced  grant ``GRBs",  by the 
I-CORE  Program of the Planning and  Budgeting Committee and The Israel Science Foundation grant No 1829/12 and by ISA grant 3-10417 .

\bibliography{Radio_Bibliography}

\appendix
\section{Binning Errors} \label{subsec:Errors}
In the following, we estimate the errors associated with the arbitrary  choice of binning, and show that they are insignificant.

To estimate this, we calculate the light-curve for different divisions of solid angles keeping the number of bins fixed at $N = 20\times20$. We enlarge and decrease randomly the size of each solid angle element and with it the number of SPH particles it contains. As we need to keep  a minimal number of SPH particles in each bin for the analysis to work, we restricted the random noise so that the smallest possible solid angle bin would still satisfy this requirement. We additionally rotated the system randomly about the $\varphi$ axis so as to begin the division at different points (essentially so that the first bin will not be constrained to the same location every time). 

\begin{figure} 
\centering
\begin{subfigure}[]{ \label{fig:t_peak_distribution_a} \epsfig{file=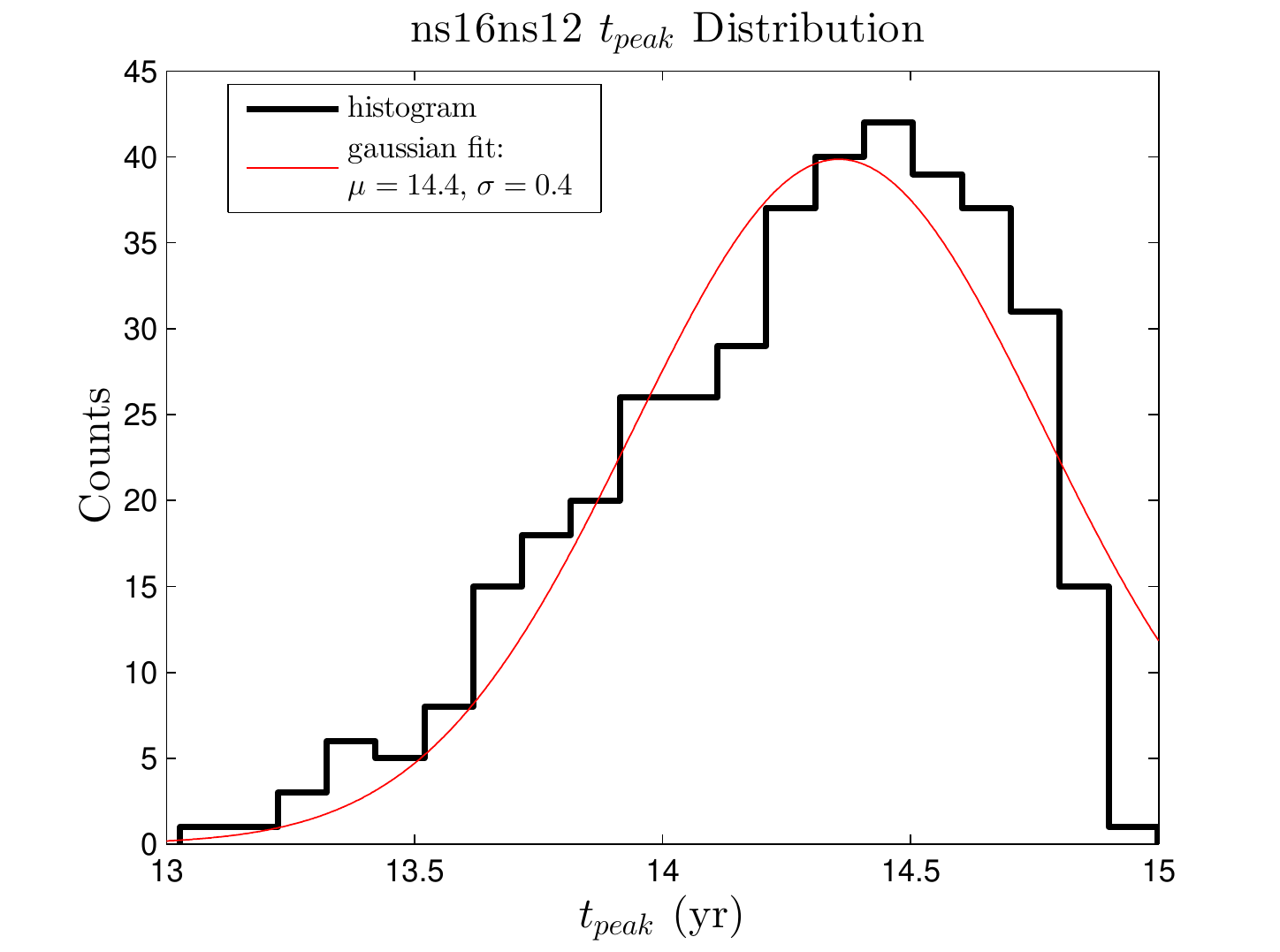,angle=0,width=0.45\textwidth} }
\end{subfigure}
~
\begin{subfigure}[]{ \label{fig:t_peak_distribution_b} \epsfig{file=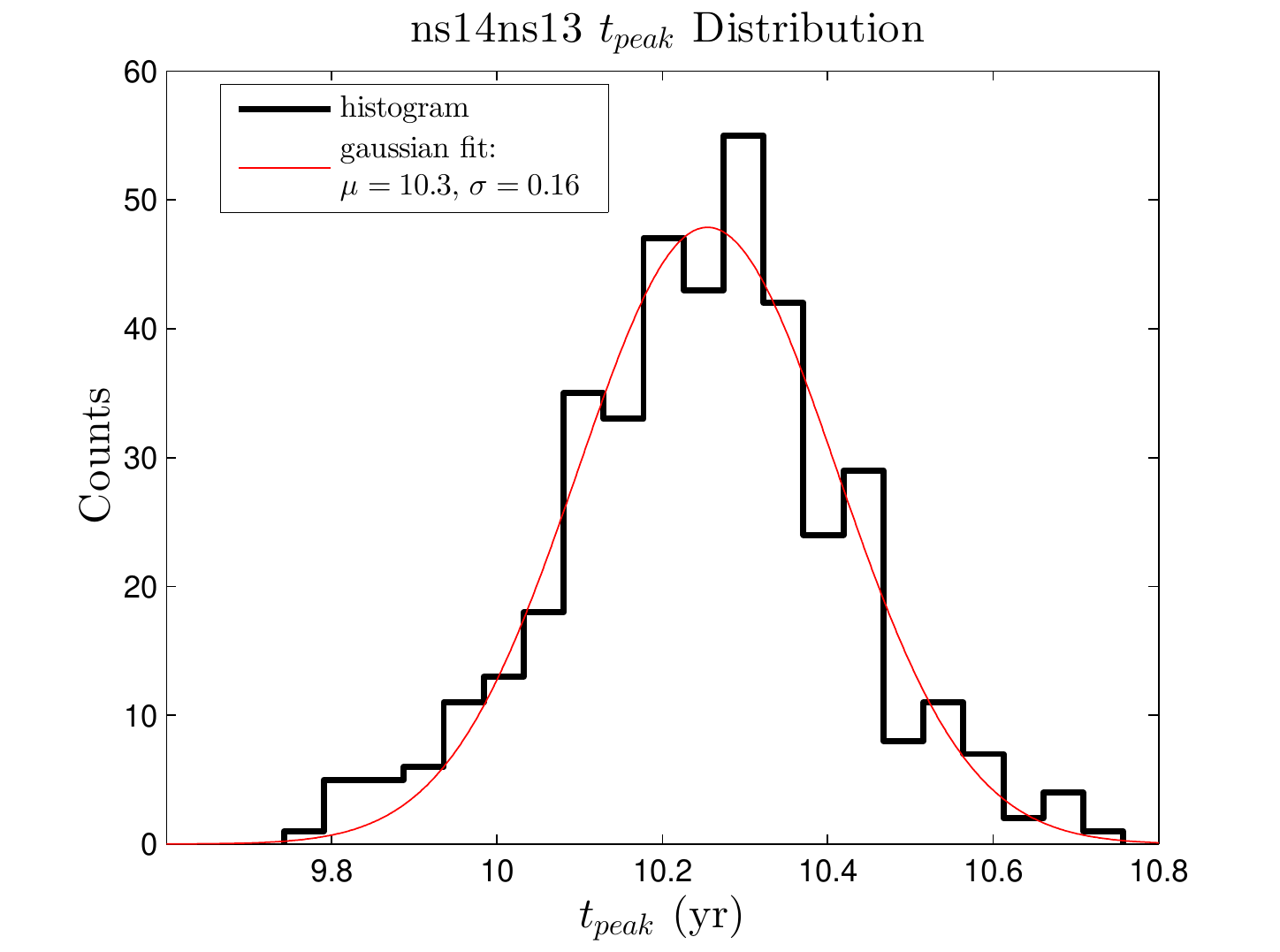,angle=0,width=0.45\textwidth} }
\end{subfigure}
~
\begin{subfigure}[]{ \label{fig:t_peak_distribution_c} \epsfig{file=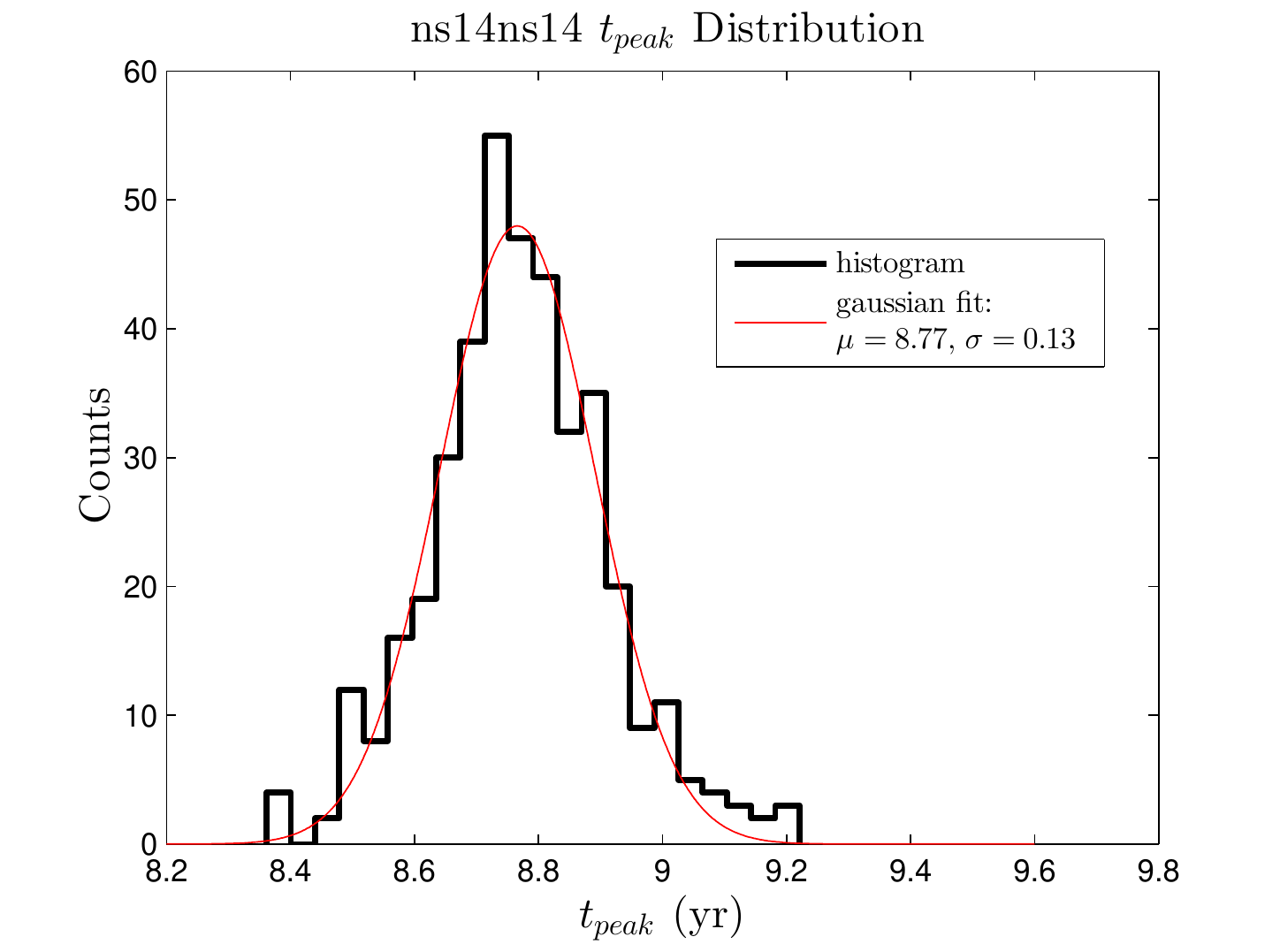,angle=0,width=0.45\textwidth} }
\end{subfigure}
\caption{Distributions of the total light-curve's peak time, $t_{peak}$, calculated using an ensemble of randomized $N = 20 \times 20$ angular divisions. This illustrates the variability in the peak time due to the arbitrary choice of solid angle bins. As this variability is of order $\sim$ 2\%, our results are consistent irrespective of any specific angular division choice. It is important to note that the stated standard deviations should not be taken as actual error bounds on our results, since the approximations of our method introduce significantly larger sources of errors than these (see \S\ \ref{subsec:Limitations_of_Our_Method}). This distribution is intended only to show that the binning method does not introduce systematic errors.} \label{fig:t_peak_distribution}
\end{figure}

Since the main variable we have studied in this work is the peak time of such light-curves, we extract this parameter and asses it's statistical variability as a measure of the errors associated with the arbitrary binning process. Fig.\ \ref{fig:t_peak_distribution} shows histograms for the peak timescale extracted from analyzing the same data using different angular discretizations. The standard deviations are of order $\sim$ 1-2\%, and therefore the binning procedure does not introduce significant systematic errors. 

Throughout the analysis we have made several assumptions and simplifications that should introduce deviations from realistic light-curves substantially larger than a few percent, and therefore it is important to use caution when considering the numerical  errors stated above.  In other words, the stated mean peak time for these runs is only an approximation of the actual peak time, and the standard deviations should not be taken as error bounds on this peak time. Nonetheless, the qualitative results such as the increase of the timescale due to the asymmetry have been established both analytically and numerically and as such the are robust and valid.

\end{document}